\documentclass[fleqn,usenatbib]{mnras}

\usepackage{newtxtext,newtxmath}

\usepackage[T1]{fontenc}

\DeclareRobustCommand{\VAN}[3]{#2}
\let\VANthebibliography\thebibliography
\def\thebibliography{\DeclareRobustCommand{\VAN}[3]{##3}\VANthebibliography}

\usepackage{graphicx}	
\usepackage{amsmath}	
\usepackage{soul}
\usepackage{pifont}
\usepackage{booktabs}
\usepackage{xcolor}

\newcommand{\fable}{{\sc fable}}%
\newcommand{\RadioWeak}{\textcolor{RadioWeakColor}{RadioWeak}}
\newcommand{\RadioStrong}{\textcolor{RadioStrongColor}{RadioStrong}}
\newcommand{\Quasar}{\textcolor{QuasarColor}{Quasar}}
\newcommand{\Fiducial}{\textcolor{FiducialColor}{Fiducial}}
\newcommand{\NoFeedback}{\textcolor{NoFeedbackColor}{NoAGN}}
\newcommand{\DMOtext}{DMO}
\newcommand{\DMO}{\textcolor{DMOColor}{\DMOtext}}

\newcommand{\linelegend}{\includegraphics[width=\columnwidth]{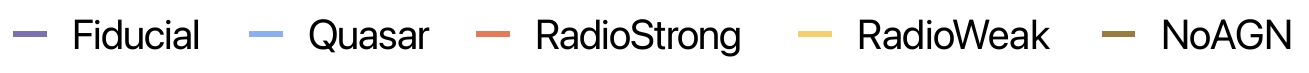}\vspace{-0.3cm}}

\newcommand{\Pmm}{P_\text{mm}}
\newcommand{\Pmf}{P_\text{mm}^{\text{M}_\text{min} < \text{M}_\text{gal}}}
\newcommand{\Pme}{P_\text{mm}^\text{excluded}}
\newcommand{\Cmfe}{\mathcal{C}_\text{filt,exc}}
\newcommand{\contrast}{\mathcal{D}}
\newcommand{\PDMO}{P_\text{mm, DMO}}
\newcommand{\Pmmhalo}{P_\text{mm, halos}}
\newcommand{\PDMOfilt}{P_\text{mm, DMO}^{\text{M}_\text{min} < \text{M}_\text{gal}}}

\newcommand{\fbar}{f_\text{baryon}}
\newcommand{\kscaleten}{k_\text{scale} = 10\,\hcMpc}
\newcommand{\kscaleone}{k_\text{scale} = 1\,\hcMpc}
\newcommand{\Mhalo}{M_\text{halo}}
\newcommand{\Mstar}{M_{*}}
\newcommand{\MBH}{M_\text{BH}}

\newcommand{\hcMpc}{\;\text{h}\;\text{cMpc}^{-1}}

\newcommand{\Msun}{\mathrm{M_{\odot}}}

\newcommand{\rowlegendfull}{\vspace{-0.1cm}
\includegraphics[width=1.2\columnwidth]{Images/linelegend.pdf}\\
\vspace{-0.3cm}
}

\title[The impact of black hole feedback on the Fable MPS]{Stirring the cosmic pot: how black hole feedback shapes the matter power spectrum in the \textsc{fable} simulations}
\author[Martin-Alvarez et al.]{
Sergio Martin-Alvarez,$^{1}$\thanks{E-mail: \href{mailto:martin-alvarez@stanford.edu}{martin-alvarez@stanford.edu}}
Vid Iršič,$^{2,3}$
Sophie Koudmani,$^{2,4,5,6,7}$
Martin A. Bourne,$^{2,4,7}$
Leah Bigwood,$^{2,4}$\newauthor
and Debora Sijacki$^{2,4}$
\\
$^{1}$Kavli Institute for Particle Astrophysics \& Cosmology (KIPAC), Stanford University, Stanford, CA 94305, USA\\
$^{2}$Kavli Institute for Cosmology (KICC), University of Cambridge, Madingley Road, Cambridge CB3 0HA, UK\\
$^{3}$Cavendish Laboratory, University of Cambridge, 19 J. J. Thomson Ave., Cambridge CB3 0HE, UK\\
$^{4}$Institute of Astronomy, University of Cambridge, Madingley Road, Cambridge CB3 0HA, UK\\
$^{5}$St Catharine's College, University of Cambridge, Trumpington Street, Cambridge CB2 1RL, UK\\
$^{6}$Center for Computational Astrophysics, Flatiron Institute, 162 5$^\text{th}$ Avenue, New York, NY 10010, USA\\
$^{7}$Centre for Astrophysics Research, Department of Physics, Astronomy and Mathematics, University of Hertfordshire, College Lane, Hatfield, AL10 9AB, UK\\
}

\date{MNRAS, accepted}

\pubyear{2025}

\begin{document}
\label{firstpage}
\pagerange{\pageref{firstpage}--\pageref{lastpage}}
\maketitle

\begin{abstract}
Understanding the impact of baryonic physics on cosmic structure formation is crucial for accurate cosmological predictions, especially as we usher in the era of large galaxy surveys with the Rubin Observatory as well as the Euclid and Roman Space Telescopes. A key process that can redistribute matter across a large range of scales is feedback from accreting supermassive black holes. How exactly these active galactic nuclei (AGN) operate from sub-parsec to Mega-parsec scales however remains largely unknown. To understand this, we investigate how different AGN feedback models in the \fable~simulation suite affect the cosmic evolution of the matter power spectrum (MPS). 
Our analysis reveals that AGN feedback significantly suppresses clustering at scales $k \sim 10 \hcMpc$, with the strongest effect at redshift $z = 0$ 
causing a reduction of $\sim 10\%$ with respect to the dark matter-only simulation. This is due to the efficient feedback in both radio (low Eddington ratio) and quasar (high Eddington ratio) modes in our fiducial \fable~model. 
We find that variations of the quasar and radio mode feedback with respect to the fiducial \fable~model have distinct effects on the MPS redshift evolution, with radio mode being more effective on larger scales and later epochs. Furthermore, MPS suppression is dominated by AGN feedback effects inside haloes at $z = 0$, while for $z \gtrsim 1$ the matter distribution both inside and outside of haloes shapes the MPS suppression. Hence, future observations probing earlier cosmic times beyond $z \sim 1$ will be instrumental in constraining the nature of AGN feedback.
\end{abstract}




\definecolor{darkgreen}{rgb}{0.0,0.6,0.0}
\definecolor{DMOColor}{rgb}{0.3,0.3,0.3}

\definecolor{FiducialColor}{rgb}{0.301961,0.262745,0.588235} 
\definecolor{QuasarColor}{rgb}{0.388235,0.584314,0.925490} 
\definecolor{RadioStrongColor}{rgb}{0.850980,0.313725,0.113725} 
\definecolor{RadioWeakColor}{rgb}{0.949020,0.745098,0.250980} 
\definecolor{NoFeedbackColor}{rgb}{0.470588,0.313725,0.000000} 

\section{Introduction}
\label{s:Introduction}
The underlying cosmology of our Universe dictates the properties and evolution of cosmic structure. One of these is the distribution of mass in our Universe, which has been mapped through both structure formation and late-time surveys \citep[e.g.,][]{Heymans2021, DES2022, Qu2024}, and early Universe observations of the Cosmic Microwave Background \citep{Hinshaw2013, PlanckCollaboration2016b}. Numerical and observational studies have shown that baryonic physics, specifically feedback processes from stars and black holes, may significantly impact the distribution of matter across the cosmic web \citep[e.g.,][]{Seljak2000,VanDaalen2011,Chisari2018,Secco2022}.
With programmes like the Vera C. Rubin Observatory \citep{LSST2019}, the Euclid Space Telescope \citep{Euclid2011_clean} and the Roman Space Telescope \citep{WFIRST2013_clean} preparing to map the Large-Scale Structure (LSS) with unprecedented accuracy, precise theoretical model predictions are urgently required to understand the processes that shape the distribution of galaxies and the underlying matter across cosmic time.

Baryonic feedback plays a crucial role in the formation of individual galaxies, where supernovae and AGN activity have been identified as some of the key processes \citep[e.g.,][]{White1991, Sijacki2007, Hopkins2014, Habouzit2017, Rosdahl2018, Trebitsch2020}. Such feedback is required to reconcile local observations with small-scale challenges to our $\Lambda$CDM model \citep{Bullock2017} as well as to produce a realistic global star formation history \citep{Madau2014}, or massive quenched elliptical galaxies and the brightest cluster galaxies \citep{McNamara2007, Fabian2012}. 
Through galactic outflows and AGN-driven winds, these feedback processes also provide a channel for galaxies to interact with their larger-scale environment and the local distribution of matter, and influence statistics such as the matter power spectrum (MPS) \citep{VanDaalen2011, Chisari2019}. Moreover, a wide range of not-so-well-understood baryonic feedback processes such as cosmic rays or magnetism are gaining popularity in the realistic modelling of outflows from galaxies \citep[e.g.,][]{Pakmor2016, Girichidis2018, Martin-Alvarez2020, Hopkins2020, Farcy2022, Beckmann2022a, Martin-Alvarez2023, Curro2024}, and may significantly affect how galaxies shape the local distribution of matter. 

AGN feedback is the main process regulating the evolution of the most massive galaxies, galaxy 
groups, and galaxy clusters \citep{Sijacki2007, Cavagnolo2010, Bourne2017, Chisari2018, Beckmann2019, BourneYang23Review}. The outflows driven by AGN are extremely energetic and can reach scales up to $\sim$Mpc, making this form of feedback the most important for the cosmic distribution of matter \citep[e.g.,][]{VanDaalen2011, Mead2015, McCarthy2018, Chisari2019}, primarily through redistribution of matter within and beyond the largest haloes \citep[e.g.,][]{vanDaalen2015, vanDaalen2020, vanLoon2023}. While the influence of AGN feedback can be captured through simple halo models \citep{Seljak2000, Mead2021}, due to the complex relationship between the small-scale regulation of accretion onto supermassive black holes (SMBHs), galaxy formation physics, and the large-scale effects of AGN feedback, cosmological simulations are required to understand its effect on the MPS. 

Multiple studies employing some of the largest and most sophisticated cosmological simulations to date \citep[e.g.,][]{Vogelsberger2014a, Hellwing2016, Springel2018, Chisari2018, vanDaalen2020, Sorini2022, vanLoon2023, Schaye2023, Gebhardt2024} have established that AGN feedback affects the MPS at scales $k \gtrsim 0.5 \hcMpc$. The resulting power suppression with respect to the dark matter non-linear prediction in these models reaches up to $\sim 20\%$ \citep{Chisari2019}. While these different simulations display similar qualitative behaviour, quantitative differences across results are significant, emerging from different feedback implementation strategies and configurations, as well as from different model resolutions and numerical solvers.

To better comprehend the discrepancies between different simulations, a more detailed understanding of how different AGN feedback models affect the MPS is required, and how this impact emerges around different galaxies and environments. AGN feedback is an inherently multi-scale phenomenon, spanning from event horizon and accretion disc scales at which the feedback (in the form of radiation, winds and jets) is produced, out to scales beyond the host galaxy itself. As such, modelling this process in full is virtually impossible within a single simulation. Instead, cosmological simulations have to employ sub-grid models that can capture the effects of AGN feedback and how it couples to baryons at resolvable scales. These models can vary in their sophistication and their made assumptions. The simplest approach is direct thermal energy injection into cells or particles close to the black hole \citep{Springel2005, Booth2009, Schaye2015, Tremmel2019}, often combined with numerically-motivated modifications, such as minimum heating temperatures \citep{Booth2009, Schaye2015, McCarthy2017}, fixed duty cycles \citep{Henden2018, Koudmani2022} or artificial prevention of radiative cooling \citep{Tremmel2017, Tremmel2019} in order to avoid over-cooling \citep[see discussions in][]{Bourne2015, Schaye2015, Crain2015, Zubovas2016}. Other models inject momentum to surrounding gas as bipolar wind or jet-like outflows \citep{Dubois2014, WeinbergerEtAl18Illustris, DaveEtAl19Simba}, with several works including separate quasar and radio mode phases that use different energy injection schemes for each \citep{Sijacki2007, Dubois2014, Sijacki2015, Henden2018, Dubois2021}. Simulations are additionally performed over a wide range of resolutions, which itself can impact the range of gas phases captured and how feedback couples to these different phases \citep[e.g.,][]{Bourne2015, Beckmann2019, Koudmani2019, Talbot2024, Hopkins2024a}. Taking this into account, as well as the use of different codes to perform cosmological simulations, model parameters are typically calibrated to match low-redshift observables such as the galaxy stellar mass function and BH scaling relations \citep{Dubois2014, Schaye2015, Sijacki2015, Pillepich2018a} meaning that different feedback models, in different codes and at different resolutions can result in comparable galaxy populations. As such it is the galaxy properties to which simulation parameters are not tuned that can be used to differentiate between models.

One such quantity is the baryon content of groups and clusters, which has been suggested as a proxy for the expected suppression in the MPS (\citealt{Semboloni2011, Semboloni2013}; \citealt{McCarthy2018, Schneider2019}; \citealt{Debackere2020, vanDaalen2020, Salcido2023}). The AGN model in the original Illustris suite of simulations was too effective at expelling gas from groups and low mass clusters \citep{Genel2014aa}, and indeed, the MPS suppression found in Illustris is more extreme than that found in other simulations that retain higher baryon fractions \citep{Chisari2019, vanDaalen2020}. The \fable~simulation suite remedied this problem by modifying the feedback models employed in Illustris, making the quasar mode more effective and the radio mode less explosive \citep{Henden2018} to achieve a better agreement to observations of group and cluster baryon content. In determining their fiducial AGN model, other variations were performed with a total of four presented in Appendix~A of \citet{Henden2018}, which result in different present-day stellar and gas fractions in groups and clusters. These variations provide an ideal testbed to study the effect of different AGN feedback models on the MPS, which provides a key motivation for the work presented here. 
 
We describe the \fable~simulations in Section~\ref{ss:Simulation}, and our procedure to MPS in Section~\ref{ss:FFT}. Our main results are explored in Section~\ref{s:Results}, where we compare various AGN feedback models (Section~\ref{ss:AGNimpact}), comparing its effect on \fable~with previous simulations (Section~\ref{ss:FableVsWorld}). We explore in more detail how feedback effects vary around galaxies under different selections (halo mass, stellar mass, and black hole mass) in Section~\ref{ss:GalaxiesCuts}. Section~\ref{ss:HaloesTrace} briefly reviews how different halo mass components trace the MPS suppression from AGN at different scales and times. In Section~\ref{s:Caveats} we review the main caveats of our study, mainly stemming from the size of the \fable~computational box employed. Finally, we conclude this manuscript in Section~\ref{s:Conclusions} with a summary of our work.

\section{Numerical methods}
\label{s:Methods}

\begin{table*}
\caption{Overview of the AGN feedback model variations, listing the Eddington fraction threshold for switching from the radio mode (RM) to the quasar mode (QM) ($f_\mathrm{Edd,QM}$), the radiative efficiency ($\epsilon_\mathrm{r}$), the quasar mode feedback efficiency ($\epsilon_\mathrm{f}$), the length of the quasar mode duty cycle ($\delta t_\mathrm{QM}$), the radio mode feedback efficiency ($\epsilon_\mathrm{m}$), and the fractional BH mass increase required for triggering a radio mode feedback event ($\delta_\mathrm{BH}$).}
\centering
\begin{tabular}{@{}llllllll@{}}
\toprule
\textbf{Name} & \textbf{Feedback}  & \textbf{Radiative} & \textbf{QM}  & \textbf{QM}  & \textbf{RM} & \textbf{RM fractional} & \textbf{Comments} \\
 & \textbf{switch}  & \textbf{efficiency} & \textbf{efficiency} & \textbf{duty cycle} & \textbf{efficiency} & \textbf{mass increase} & \\
 & $f_\mathrm{Edd,QM}$  & $\epsilon_\mathrm{r}$ & $\epsilon_\mathrm{f}$ & $\delta t_\mathrm{QM}$ [Myr] & $\epsilon_\mathrm{m}$ & $\delta_\mathrm{BH}$ &  \\\midrule
\Fiducial~(\textit{QuasarDutyRadioStrong}) & 0.01 & 0.1 & 0.1 & 25 & 0.8 & 0.01 & standard \fable \ set-up \\
\RadioWeak~(\textit{NoDutyRadioWeak}) & 0.05 & 0.1 & 0.1 & - & 0.4 & 0.001 & no QM duty, weak RM \\
\RadioStrong~ (\textit{NoDutyRadioStrong}) & 0.05 & 0.1 & 0.1 & - & 0.8 & 0.01 & no QM duty, strong RM \\
\Quasar~ (\textit{QuasarDutyRadioWeak}) & 0.05 & 0.1 & 0.1 & 25 & 0.4 & 0.001 & QM duty, weak RM \\
\NoFeedback~ (\textit{NoAGNFeedback}) & - & - & - & - & - & - & no AGN feedback \\
\midrule
\DMO~ (\textit{Dark matter only}) & - & - & - & - & - & - & no baryons \\
\midrule
\textit{Illustris} & 0.05 & 0.2 & 0.05 & - & 0.35 & 0.15 & Illustris set-up for reference \\
 \bottomrule
\end{tabular}
\label{tab:sims_overview}
\end{table*}

\subsection{\fable \ simulations}
\label{ss:Simulation}
In this section, we provide a brief summary of the \fable \ simulation suite \citep{Henden2018,Henden2019,Henden2020}, which we employ for our investigation into the impact of AGN feedback models on the MPS and the galaxy bias. For a detailed description of the \fable~set-up and the calibration of the simulations see \citet{Henden2018}.

\subsubsection{Basic simulation properties}

The \fable \ simulations were performed with the \textsc{arepo} code \citep{Springel2010aa}, where the equations of hydrodynamics are solved on a moving unstructured mesh defined by the Voronoi tessellation of a set of discrete points which (approximately) move with the velocity of the local flow. The gravitational interactions are modelled via the TreePM method with stars and DM represented by collisionless particles.

The \fable \ simulation suite comprises cosmological volumes as well as zoom-in simulations of groups and clusters. Here we focus on the cosmological volume simulations to investigate the clustering of matter at large scales (rather than examining individual haloes). These 40 $h^{-1} \, \mathrm{Mpc} $ ($h = 0.679$) boxes are evolved using initial conditions for a uniformly sampled cosmological volume based on the Planck cosmology \citep{PlanckCollaboration2016} with $512^3$ DM particles, yielding a resolution of $m_\mathrm{DM} = 3.4 \times 10^7 ~ h^{-1}\, \Msun$, and initially $512^3$ gas elements with target gas mass resolution $\overline{m}_\mathrm{gas} = 6.4 \times 10^{6} ~ h^{-1}\, \Msun$. The gravitational softening is set to $2.393 ~ h^{-1} \, \mathrm{kpc}$ in physical coordinates below $z=5$ and held fixed in comoving coordinates at higher redshifts. Notably, this leads to a suite of high-resolution simulations, albeit with a comparatively small cosmological volume when addressing cosmological stastics such as the MPS. We discuss the main caveats resulting from this in Section~\ref{s:Caveats}.

The \fable \ galaxy formation model is based on Illustris \citep{Vogelsberger2013a,Vogelsberger2014,Genel2014aa,Torrey2013,Sijacki2015}, with the prescriptions for radiative cooling \citep{Katz1996,Wiersma2009a}, uniform ultraviolet background \citep{Faucher-Giguere2009}, chemical enrichment \citep{Wiersma2009} and star formation \citep{Springel2003aa} unchanged from the Illustris model. The stellar and AGN feedback prescriptions, on the other hand, are modified to improve agreement with the present-day galaxy stellar mass function and to match the gas mass fractions in observed massive haloes.

\subsubsection{Stellar feedback}
In the Illustris galactic wind model \citep{Vogelsberger2013a}, wind particles are launched from star-forming regions driven by the available energy from core-collapse SNe.

This model is also adopted in \fable \ with a few modifications to the parameters that govern the wind energetics. Specifically, the wind energy factor $\epsilon_\mathrm{W,SN}$, which gives the fraction of energy available from each core collapse supernova, is increased to $\epsilon_\mathrm{W,SN} = 1.5$ in \fable \ compared to the Illustris value of $\epsilon_\mathrm{W,SN} = 1.09$. Furthermore, one-third of the wind energy is injected as thermal energy in \fable, whilst in Illustris the stellar-feedback-driven winds are purely kinetic. Overall, this leads to more energetic stellar feedback which more efficiently dissipates the released energy to the gas, and somewhat more effectively regulating star formation in low-mass haloes (see \citealt{Henden2018} for details; the same method is used by \citealt{Marinacci2014}).

\subsubsection{Black hole seeding and growth}
BHs are modelled as collisionless particles and are seeded into DM haloes above a mass threshold of $5 \times 10^{10} \ h^{-1} \, \Msun$ with a seed mass of $M_\mathrm{BH,seed}= 10^{5} \ h^{-1}\, \Msun$.

Subsequently, these BHs may grow via BH -- BH mergers and gas accretion following the Eddington-limited Bondi-Hoyle-Lyttleton accretion rate with boost factor $\alpha = 100$ \citep{Hoyle1939,Bondi1944,Springel2005}. For all AGN models the radiative efficiency is set to a constant $\epsilon_\mathrm{r} = 0.1$ and $(1 - \epsilon_\mathrm{r})$ of the accreted mass is added to the BH particle mass at each timestep.

Lastly, we note that the BHs are pinned to the potential minimum of their host halo to prevent spurious BH movement due to numerical heating \citep[see][for details on the BH seeding and growth models]{Sijacki2007, Vogelsberger2013a}.

\subsubsection{AGN feedback} 
\label{sss:AGNFeedbackMethods}
Analogously to Illustris, the AGN feedback in \fable~is based on a two-mode model, with the quasar mode operating at high Eddington ratios \citep[see][]{DiMatteo2005,Springel2005} and the radio mode being activated at low Eddington ratios \citep[see][]{Sijacki2007}. For the fiducial \fable~simulation set-up, this switch occurs at an Eddington ratio of $f_\mathrm{Edd,QM} = 0.01$ (compared to $f_\mathrm{Edd,QM} = 0.05$ in Illustris).

In the quasar mode, a fraction $\epsilon_\mathrm{f}=0.1$ of the AGN luminosity is isotropically injected as thermal energy. In Illustris, this thermal energy injection happens continuously, which can lead to artificial overcooling as small amounts of energy are distributed preferentially into the densest material over a large gas mass due to the limited gas mass resolution. In \fable, this issue is alleviated by introducing a duty cycle with an approach similar to that of \citet{Booth2009}, whereby thermal energy is accumulated over $\delta t_\mathrm{QM} = 25$~Myr before being released in a single event, allowing high feedback temperatures, and hence longer cooling times, to be reached. Such a feedback cycle is also, at least qualitatively, consistent with episodic accretion observed in high-resolution simulations \citep{Ciotti2010, Torrey2017, Costa2018b}.

In the radio mode, the feedback energy is coupled to the gas as hot buoyantly-rising bubbles to mimic those inflated by jets \citep{McNamara2007, Fabian2012, BourneYang23Review}, with the duty cycle of these bubble injections set by the fractional BH mass growth
$\delta_\mathrm{BH} = \delta M_\mathrm{BH} / M_\mathrm{BH}$.
In \fable, this threshold is set to $\delta_\mathrm{BH} = 0.01$ -- much smaller than the Illustris value of $\delta_\mathrm{BH} = 0.15$. The bubble energy content is determined as $\epsilon_\mathrm{m} \epsilon_\mathrm{r} \mathrm{c^{2}} \delta M_\mathrm{BH}$ with the radio mode coupling efficiency set to $\epsilon_\mathrm{m} = 0.8$ in the fiducial \fable \ model. This yields a similar effective radio mode efficiency $\epsilon_\mathrm{m} \epsilon_\mathrm{r} = 0.08$ as in the Illustris model (where the effective radio mode efficiency is set to 7 percent). The lower $\delta_\mathrm{BH}$ then results in more frequent and less energetic bubbles in \fable \ compared to the Illustris set-up.

\subsubsection{AGN model variations}

In addition to the fiducial \fable \ model, \citet{Henden2018} explore three additional AGN feedback parametrizations:

\begin{itemize}
    \item[--] The \RadioStrong~set-up, which has the same radio mode parameters as the fiducial run but no quasar duty cycle.
    \item[--] The \Quasar~set-up, which employs a quasar duty cycle but has significantly weaker radio mode feedback with a lower threshold for bubble injections ($\delta_\mathrm{BH} = 0.001$) and a lower coupling efficiency ($\epsilon_\mathrm{m} = 0.4$).
    \item[--] The \RadioWeak~set-up, which does not have a quasar duty cycle \textit{and} employs the weaker radio mode feedback.
\end{itemize}

Together with the fiducial run, these three alternative AGN set-ups then allow us to isolate the impact of the quasar duty cycle and increasing the strength of the radio mode feedback. Note that all of the additional runs also have a higher critical Eddington fraction for switching to the quasar mode ($f_\mathrm{Edd,QM} = 0.05$, as in Illustris).

Furthermore, we also analyse the results from an additional \fable \ model variation, \NoFeedback, which was performed without seeding any black holes, therefore providing a useful reference run without any AGN feedback.

The four AGN runs and the no-AGN run form the core of our analysis and the corresponding AGN parameters for these five set-ups are listed in Table~\ref{tab:sims_overview}. For reference, the corresponding parameters for the original Illustris simulation set-up are also given.

\subsubsection{Halo and galaxy identification}
\label{ss:Cuts}
For our analysis, we identify DM haloes (`groups') and galaxies (`subhaloes') via the friends-of-friends (FoF) and \textsc{subfind} algorithms \citep{Davis1985, Springel2001aa, Dolag2009b}, respectively. The FoF search linking length is set to 0.2 times the mean particle separation. Within the FoF groups gravitationally bound systems are identified as subhaloes, as found by \textsc{subfind}. The central subhalo corresponds to the subhalo at the minimum potential of the FoF group whilst all other subhaloes in the same group are categorised as satellites.

We characterise the galaxy properties employing the total stellar mass of each subhalo as the stellar mass of each galaxy $M_*$, the central black hole mass in each subhalo as $\MBH$, and the subhalo total mass as $\Mhalo$. For the \NoFeedback~and \DMO~models, we estimate the $\MBH$ and $\Mstar$ of each galaxy interpolating its $\Mhalo$ in the Fiducial model relationships. These relationships are presented and described in Section~\ref{ss:GalaxiesCuts}. 

\subsection{Power spectra}
\label{ss:FFT}

We focus our investigation on the matter power spectrum, which provides information of the matter clustering at different scales, studied here in Fourier space and characterised by a wavelength $k$ in units of $\hcMpc$. In order to extract from the \fable~ simulations the studied matter power spectra and cross-correlations between quantities, we make use of the {\sc FFTW} library\footnote{The FFTW library can be found at \href{http://www.fftw.org/}{http://www.fftw.org/}.}. We project the entire computational domain of the simulation onto a uniform grid of $1024^3$ cells. Hence the computational domain with a physical size of $L_\text{FFT} \sim 40\,\text{h}^{-1}\,\text{cMpc}$ ($59\,\text{cMpc}$) is resolved down to $dx_\text{FFT} \sim 39 \,\text{h}^{-1}\,\text{ckpc}$ ($58\,\text{ckpc}$). Consequently, each of our spectra spans from $k_\text{min} \sim 0.2\hcMpc$ to $k_\text{max} \sim 74\hcMpc$, although we note that our results are affected by the limited simulation volume, outlined in Section~\ref{s:Caveats}, particularly at $k_\text{max} \lesssim 1\hcMpc$. To obtain the MPS, we project onto a 3D grid all the particles included in \fable~ (i.e. dark matter, stars, gas and black holes), employing a simple nearest grid point (NGP) interpolation. 
Finally, whenever performing a cross-correlation between two scalar fields, we compute this through their multiplication in Fourier space, finally averaging onto a 1D k-space binning.

\section{Results}
\label{s:Results}

\subsection{A qualitative comparison of AGN impact around massive galaxies}
\label{ss:visualization}

\begin{figure*}
    \centering
    \includegraphics[width=1.89\columnwidth]{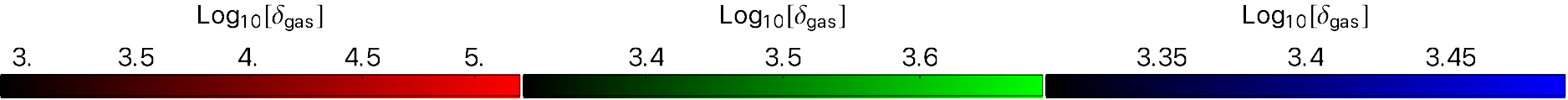}\\ 
    \vspace{-0.1cm}
    \includegraphics[width=2.1\columnwidth]{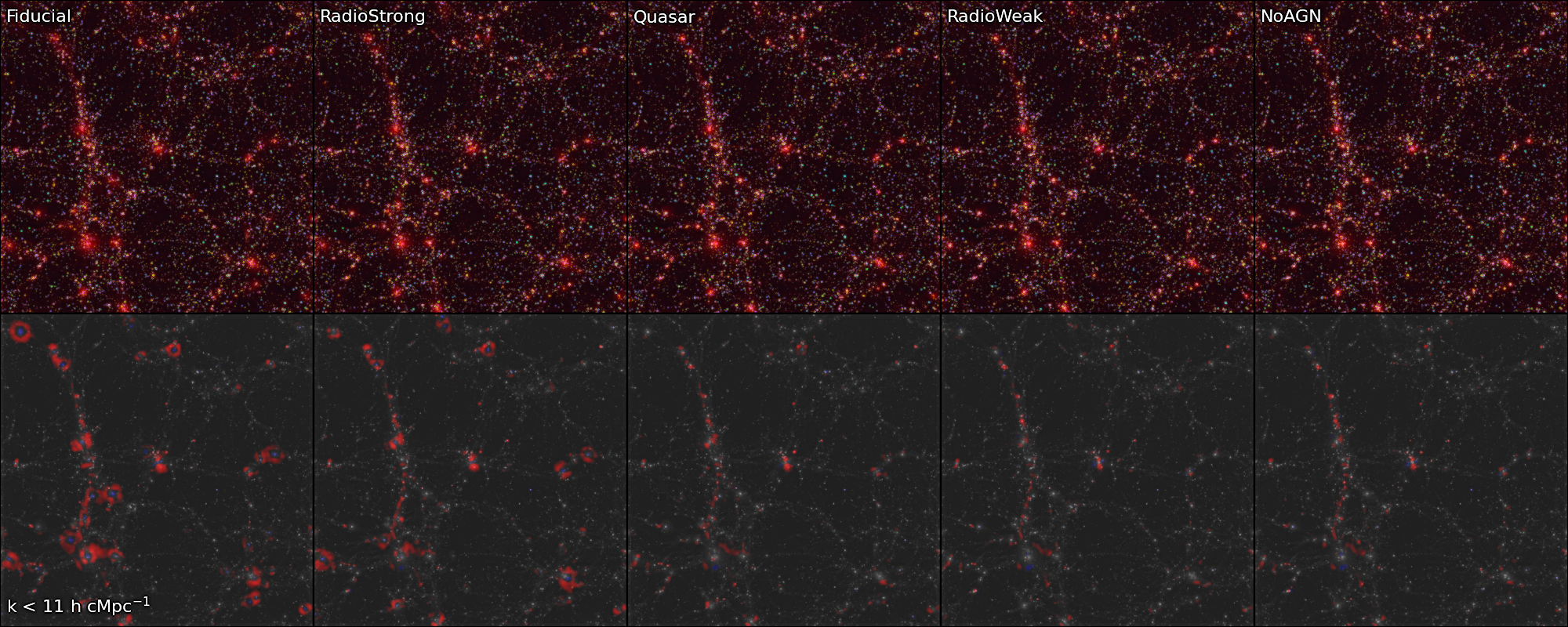}\\
    \vspace{-0.1cm}
    \includegraphics[width=1.26\columnwidth]{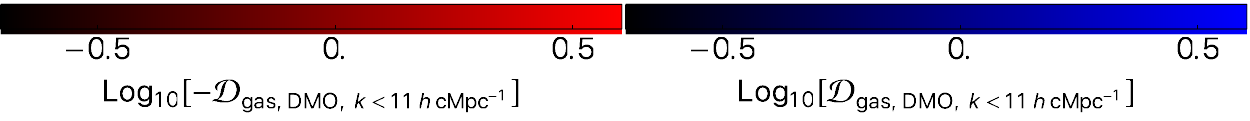}\\ 
    \vspace{-0.25cm}
    \caption{{\bf(Top row)} Gas mass density contrast projections of the full \fable~simulated domain, with mass segregated into large scales (red; $k < 11 \hcMpc$), intermediate scales (green; $11 \hcMpc < k < 22 \hcMpc$) and small scales (blue; $22 \hcMpc < k$) employing Fourier space filtering (see text). 
    {\bf(Bottom row)} Relative matter power contrast between the gas overdensity of a given simulation and the total mass overdensity of the \DMO~model, $\contrast_\text{gas, DMO}$. Relative power suppression and enhancement are shown in red and blue, respectively. We include a depiction of the overdensity from the top row using a gray scale for visual guidance. We apply a Fourier scale filtering to segregate large scales ($k < 11 \hcMpc$). We observe ring-like structures where significant power suppression (red circular shapes) occurs around massive galaxies. These rings are especially prominent for the strongest feedback models (\Fiducial~and \RadioStrong; first and second columns), and clearly absent from the \NoFeedback~simulation (rightmost column; no AGN feedback).}
    \label{fig:ScalesView}
\end{figure*}

As the hot gas ejected by AGN feedback escapes from galaxies and expands against the circumgalactic and intergalactic medium, it leads to the ejection of gas from the densest regions of the cosmic web, reducing the amount of clustering on the smallest scales. To provide a qualitative visualization of this effect, we display in Figure~\ref{fig:ScalesView} overdensity and density contrast maps, with each of the studied simulations corresponding to a different column. The first row of panels shows RGB projections where the overdensity at different scales is represented in colours: large (red; $k < 11 \hcMpc$), intermediate (green; $11 \hcMpc < k < 22 \hcMpc$) and small (blue; $k > 22 \hcMpc$). Such separation is generated in 3D Fourier space, through a band-pass filter that isolates a specific range of scales applied to the entire simulated domain. The resulting remaining power is then converted back to configuration space, and the corresponding overdensity field employed to generate the projections. As the baryonic impact amounts to proportions of no more than $\sim20\%$ in most simulations \citep[e.g., the set compiled by][]{Chisari2019}, the overdensity projections in this first row show only subtle differences across simulations. 

To explore the impact of different AGN feedback models, we focus on relative changes between simulations, which are normalised with respect to the dark matter only (\DMO) \fable~model (Figure~\ref{fig:ScalesView})
. Accordingly, the bottom row of Figure~\ref{fig:ScalesView} shows the overdensity contrast for the gas mass with respect to the \DMO~model, $\mathcal{D}_\text{gas, DMO}$, calculated as:
\begin{equation}
\contrast_\text{component, ref}=\frac{\delta_\text{component, sim} - \delta_\text{ref}}{\delta_\text{ref}}\,,
\label{eq:contrast}
\end{equation}
where $\delta_\text{component}$ is the overdensity of a given component, computed for a given simulation {\it sim} with respect to a reference model {\it ref}. In Figure~\ref{fig:ScalesView}, positive values (shown in blue) indicate that the baryonic gas has a higher overdensity than the (total) overdensity of the \DMO~case, whereas negative values (shown in red) indicate a lower overdensity instead. The colour scales are fixed equally for all panels in each row, and we separate large scales in the bottom row, following the Fourier low-k-pass filter method outlined above. We include the underlying gas overdensity distribution in gray to guide the eye. All models show some increase of power with respect to the \DMO~simulation within densest nodes of the cosmic web. Such denser structures are primarily driven by baryonic cooling. The \NoFeedback~panel (rightmost column; note that \NoFeedback~does not include AGN feedback but still has supernova stellar feedback), illustrates how baryonic cooling and SN feedback suffice to drive some local mild power suppression with respect to the \DMO~scenario. However, models with strong AGN feedback show clear circular red structures around massive galaxies. These are associated with a considerable reduction of power at large scales, driven by AGN activity evacuating matter towards larger radii. The \Quasar~model appears to mostly enhance events of bi-directional power suppression. By combining the duty cycle (\Quasar) and increased radio mode strength (\RadioStrong) modifications to the AGN model, the \Fiducial~simulation has an enhanced suppression of power, where both the characteristic large-scale ring-like structures from a stronger AGN and the bi-channel ejection of the duty-cycle are intensified. We note that the isotropic or anisotropic impact of the feedback is also driven by the environment impacted, the scales reached by the AGN feedback, and even the redshift when the effect takes place (see Section~\ref{ss:HaloesTrace}).

Overall, the presence of ring-like structures for the efficient AGN models in such large-scale projections (e.g., central object in top panels), illustrates how AGN feedback is responsible for re-shaping the distribution of matter around the clusters and galaxies of the cosmic web. These structures suggest an approximately isotropic displacement of gas for the strongest AGN models (\Fiducial~and \RadioStrong), and more anisotropic effects for the duty-cycle model (\Quasar). We will show below how simulations with these spherical matter ejections display a larger power suppression at intermediate cosmological scales, whereas our duty cycle model is more efficient in driving small-scale effects, and has a more complex redshift evolution. The lack of significant power suppression in the \NoFeedback~and \RadioWeak~models suggests that in the absence of efficient AGN models, other baryonic physics may only have marginal effects.

\subsection{The impact of baryons on the matter power spectrum of the \fable~simulations}
\label{ss:AGNimpact}

\begin{figure}
    \centering
    \includegraphics[width=\columnwidth]{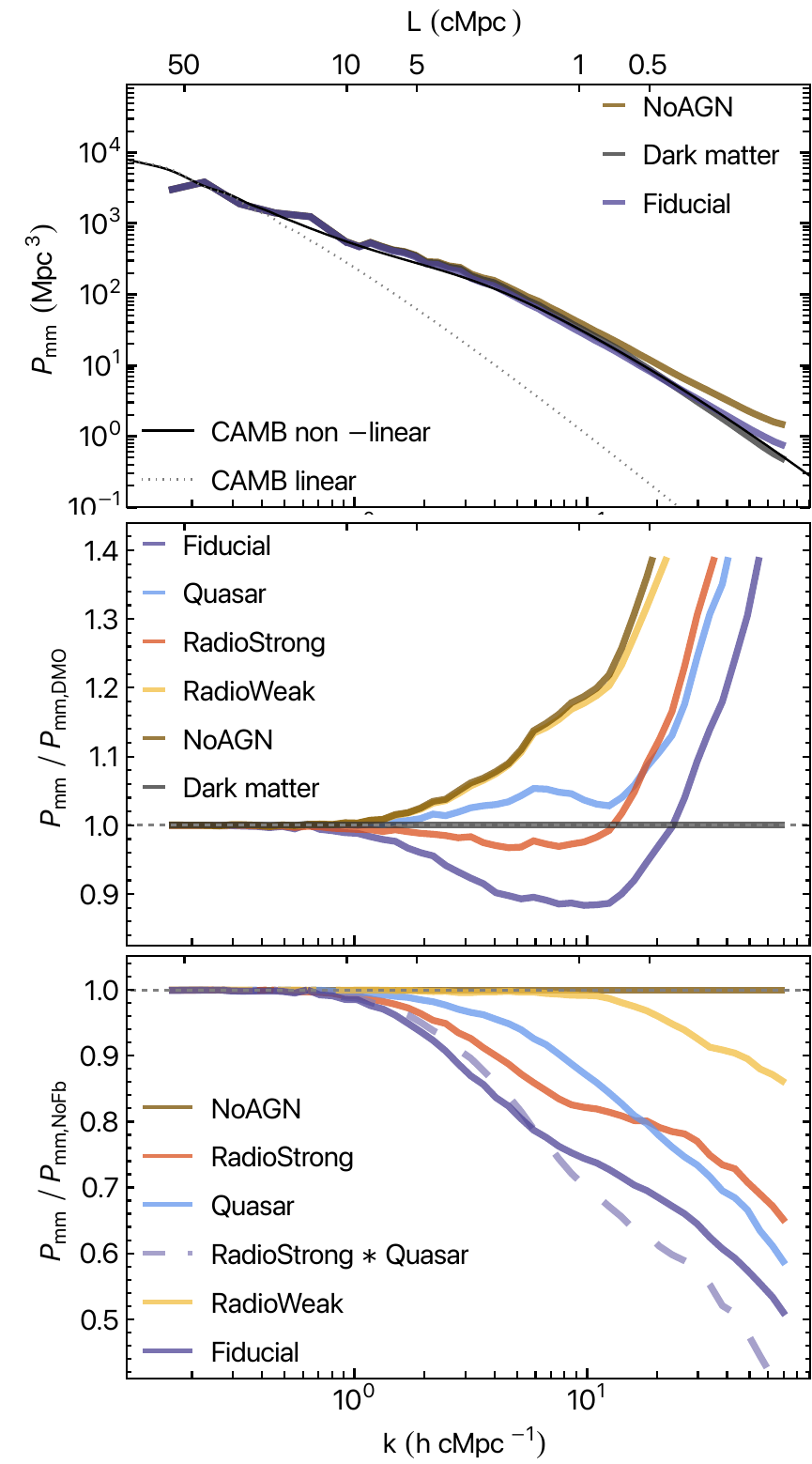}\\
    \caption{{\bf(Top panel)} Matter power spectra for the \Fiducial, \DMO~and \NoFeedback~\fable~simulations. Gray lines show the {\sc camb} non-linear (solid) and linear (dotted) predictions for the MPS adopting the same cosmology as in \fable. {\bf(Central panel)} Fractional impact of baryonic physics on the MPS at $z = 0$ obtained as the ratio of the spectrum for each model to that of the \DMO~simulation. The largest power suppression is seen in the \Fiducial~model, with the \Quasar~ and \RadioStrong~models showing an interesting crossing in relative power at scales of $k \sim 12 \hcMpc$. This illustrates how the \Quasar~duty cycle is particularly efficient in small-scale power suppression ($k \sim 10 \hcMpc$), whereas the radio mode suppresses power preferentially at $k \sim 5 \hcMpc$. {\bf(Bottom panel)} Same as the central panel, now displaying ratios to the \NoFeedback~model. We include an additional line depicting how the maximised combination of the \RadioStrong~and \Quasar~models over \RadioWeak~(see text for details) adds up, and compares with the \Fiducial~model. Separately adding up the quasar duty cycle and increased radio mode strength power suppression is approximately equivalent to their combined effect in the \Fiducial~simulation for $k \lesssim 10\,\hcMpc$. Public access to the raw \fable~MPS data for all our models is provided in the \hyperref[s:Public_Data]{Data Availability} Section.}
    \label{fig:PkRatioDM}
\end{figure}

The main statistic of interest to understand how baryonic feedback affects the distribution of matter is the MPS, which we compute as described in Section~\ref{ss:FFT}. The resulting MPS for some representative \fable~models at $z = 0$ are presented in the top panel of Figure~\ref{fig:PkRatioDM}, which includes the {\sc camb} predictions for this redshift, employing the Boltzmann solver (with $k_\text{h min} = 10^{-4} \,\hcMpc$, $k_\text{h max} = 7 \cdot 10^{1} \,\hcMpc$, using 1000 points) and for the employed \fable~cosmology (see Section~\ref{ss:Simulation}). For the non-linear prediction, we employ the Halofit model version by \citet{Mead2021}. The full-physics \fable~MPS significantly deviates from the non-linear prediction at scales $k \gtrsim 30 \hcMpc$ due to clustering from the collapse of baryons into haloes.

The impact of baryonic feedback on the MPS is typically concentrated on relatively small cosmological scales $k \sim 10\,\hcMpc$ \citep{Chisari2019}. In order to study such impact in \fable~in more detail, we show in Figure~\ref{fig:PkRatioDM} the $z = 0$ ratio of the different studied models with respect to the \DMO~case (central panel) and with respect to the \NoFeedback~model (bottom panel). We find the baryonic feedback in \fable~to only have a significant effect for scales $k > 1 \hcMpc$, (see also Section~\ref{s:Caveats}). The clustering effect of baryons dominates for scales $k \gtrsim 30 \hcMpc$ in all our models. Amongst all the simulations with AGN feedback, the \RadioWeak~model shows the lowest suppression of power, with an MPS almost equivalent to that of the \NoFeedback~model. The \Quasar~and \RadioStrong~models both significantly suppress the amount of baryonic clustering. This leads to a large deviation from the \NoFeedback~simulation, reaching a MPS comparable to the \DMO~case at scales $k \lesssim 20\,\hcMpc$. These two models also show an interesting power cross-over at scales of approximately $k \sim 12 \hcMpc$, where the \Quasar~model yields a lower power suppression at scales larger than this cross-over. The AGN duty cycle in this model concentrates feedback into periodic bursts, and AGN activity appears particularly efficient at scales $k \gtrsim 5 \hcMpc$. On the other hand, increasing the radio mode feedback strength (\RadioStrong~model) smoothly suppresses the MPS on scales of $k \sim 2 - 10 \hcMpc$. When both a stronger radio mode and quasar duty cycle are combined in the \Fiducial~model, a maximal power suppression of $\sim10\%$ below the \DMO~scenario is reached, with the suppression peaking at scales $k \sim 10 \hcMpc$. As expected, the largest relative difference in power between two models with AGN feedback at this suppression peak occurs between the \RadioWeak~and \Fiducial. 

Interestingly, the \Fiducial~power suppression has features characteristic from both models, such as the plateau in power for $k \in \left[3, 7\right]\hcMpc$ from \RadioStrong~and a dip in power at $k \sim 15\,\hcMpc$ from \Quasar. To understand whether the modifications with respect to \RadioWeak~in each of the two models can be directly combined to recover the suppression observed in \Fiducial, we include in the bottom panel of Figure~\ref{fig:PkRatioDM} a line for \RadioStrong $*$\Quasar~(pink dashed line). This is computed by removing from the \RadioWeak~$\Pmm$ the differences in power between both \RadioStrong~and \RadioWeak, and \Quasar~and \RadioWeak. Overall, `\RadioStrong $*$\Quasar' traces the \Fiducial~model well, with only a slight underestimation of the suppression towards large scales ($k < 5\,\hcMpc$), and a suppression overestimate at $k \gtrsim 10\,\hcMpc$. This hints towards an independent impact of applying a quasar duty cycle and increasing the radio mode efficiency, at least within the variation studied in \fable. We will show below that this is due to the two modes being active and effective at different redshifts. 

\begin{figure*}
    \centering
    \includegraphics[width=2.1\columnwidth]{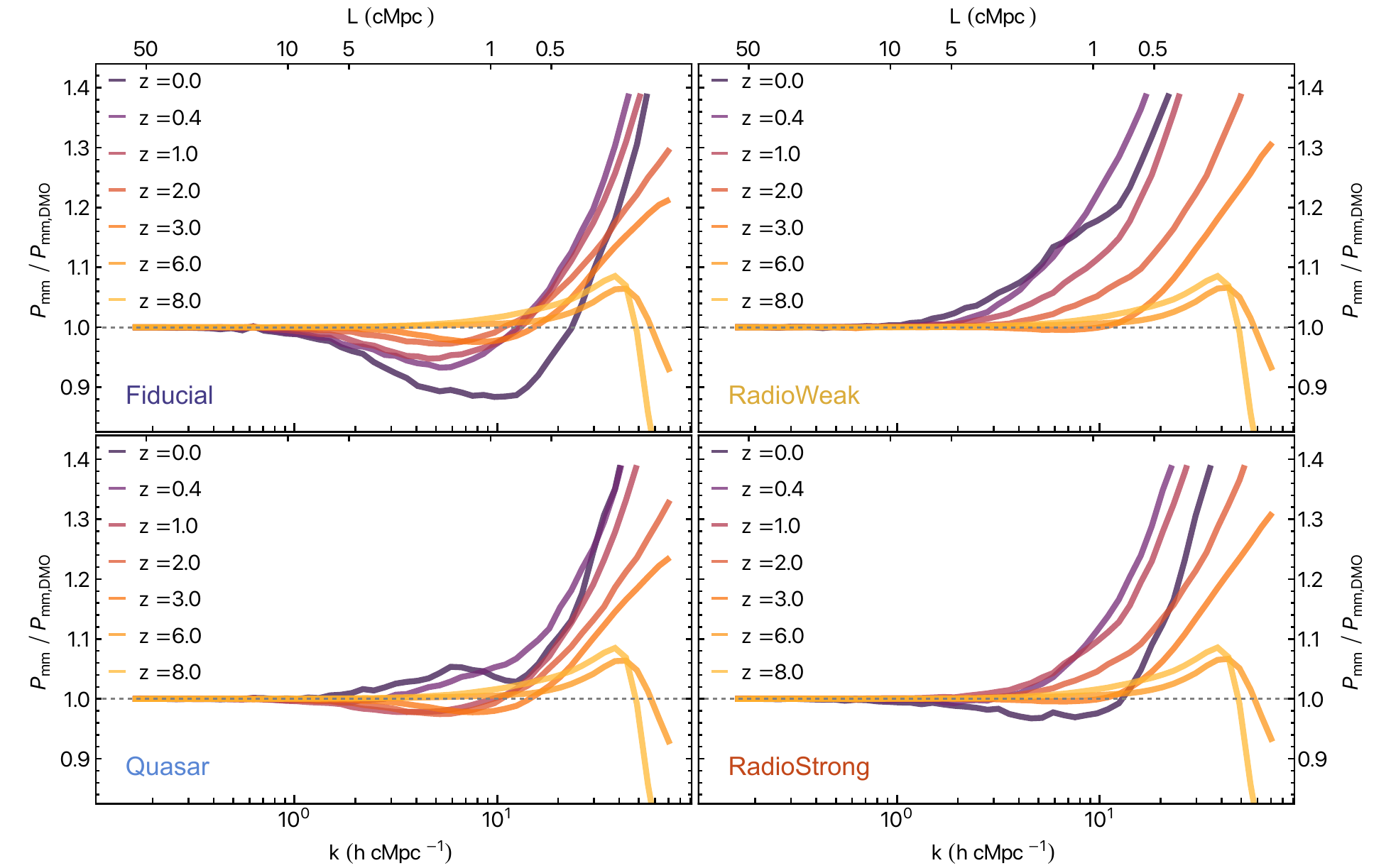}%
    \caption{Redshift evolution of the fractional impact of baryonic physics on the MPS for the \Fiducial~(top left), \RadioWeak~(top right), \Quasar~(bottom left) and \RadioStrong~(bottom right) feedback models, respectively. The \Quasar~model is more efficient at higher redshift whereas the \RadioStrong~AGN model drives a rapid power suppression at late times. Combined in the \Fiducial~simulation, these modifications to the AGN model lead to an early suppression build-up followed by an efficient decrease in relative power at low redshift. The evolution of the impact of AGN feedback on the MPS with redshift is complex, particularly after $z \sim 1$ and especially for the \Fiducial~simulation. Public access to the raw \fable~MPS data for all our models from $z = 0.0$ to $z = 2.0$ is provided in the \hyperref[s:Public_Data]{Data Availability} Section.}
    \label{fig:redshiftEvolution}
\end{figure*}

To better understand how AGN feedback, through the two separate modes, progressively carves its impact on the MPS, Figure~\ref{fig:redshiftEvolution} displays the redshift evolution of this quantity between $z = 8$ and $z = 0$ in the four studied AGN models. At high redshifts, $\Pmm$ remains similar to $\PDMO$, but as the simulations evolve, clustering is increased at small scales and AGN feedback progressively leads to power suppression at the intermediate scales. In the two simulations without the quasar duty cycle (i.e., \RadioWeak~and \RadioStrong; right column) baryonic cooling leads to higher clustering than in the \DMO~case down to $z \sim 0.5$, with the models being comparable at $z \gtrsim 1$. After this redshift, the \RadioStrong~model undergoes a significant power suppression, particularly prominent during the $z \in \left(0.5, 0.0\right]$ interval. The \RadioWeak~simulation evolves to closely resemble the \NoFeedback~case at $z = 0$, with its maximal relative deviation from it at $z \sim 0.5$, and with only a mild suppression afterwards. 

On the other hand, both models with the quasar duty cycle (i.e., \Quasar~and \Fiducial; left column) display an early suppression of power from $z \sim 3$ onwards. The early impact takes place at larger comoving scales and progressively shifts towards smaller scales as the simulation evolves, to eventually reach the peak of suppression observed at $k \sim 10 \hcMpc$ at $z = 0$. Despite this, the AGN feedback in the \Quasar~model is unable to maintain suppression over the \DMO~case after $z \sim 1$, and develops the noticeable peaks in power ($k \sim 7 \hcMpc$) and suppression ($k \sim 10 \hcMpc$) after $z \sim 0.5$. The \Fiducial~model continuously builds up relative suppression with respect to the \DMO~model during the $z \sim 3$ to $z = 0$ interval, with most of the deviation taking place after $z \sim 0.5$ as observed in the \RadioStrong~case.

The differences in the redshift evolution of the matter power spectrum for these different models are primarily driven by the temporal evolution of the quasar mode and radio mode fractions (AGN mode fractions are shown in Figure~\ref{fig:AGNmodeRedshift}, discussed in Appendix~\ref{appsec:AGN_mode_evol}), with the quasar mode dominating at high redshifts and the radio mode becoming increasingly more important towards low redshift. The quasar duty cycle in combination with strong radio mode feedback, as in the \Fiducial~model, then ensures efficient AGN feedback injection throughout cosmic history. Consequently, the effective quasar mode will lead to an earlier power suppression ($z > 1$) whereas the radio mode will be important at late times ($z < 2$). These introduce an evolutionary degeneracy that should be addressed as upcoming observatories such as the Simons Observatory probe $z \gtrsim 1$ \citep{Ade2019}.

Along these considerations, while matter clustering in the \RadioWeak~model is always above the \DMO~case, and comparable to the \NoFeedback~simulation, all the other models have $\Pmm / \PDMO < 1$ at some point during their evolution in the $3 \geq z \geq 0$ interval. The largest power suppression for both \Fiducial~and \RadioStrong~models takes place at $z = 0$, whereas it takes place during $z \sim 2 - 1$ for the \Quasar~simulation. When comparing the strongest and weakest AGN feedback models (top row; \Fiducial~vs \RadioWeak) their relative suppression of power at $k \sim 10 \hcMpc$ is of the order of $40\%$, and primarily develops at $z \lesssim 1$. The bottom row illustrates how, at small scales ($k > 10 \hcMpc$) and after $z \sim 3$, the \Quasar~AGN model has a considerably higher suppression of power than the \RadioStrong~model (bottom row). Finally, all AGN simulations, excepts perhaps \RadioWeak, have a complex evolution of their MPS with respect to the \DMO~case.

\subsection{Comparing \fable~with other simulations}
\label{ss:FableVsWorld}

\begin{figure*}
    \centering
    \includegraphics[width=2.1\columnwidth]{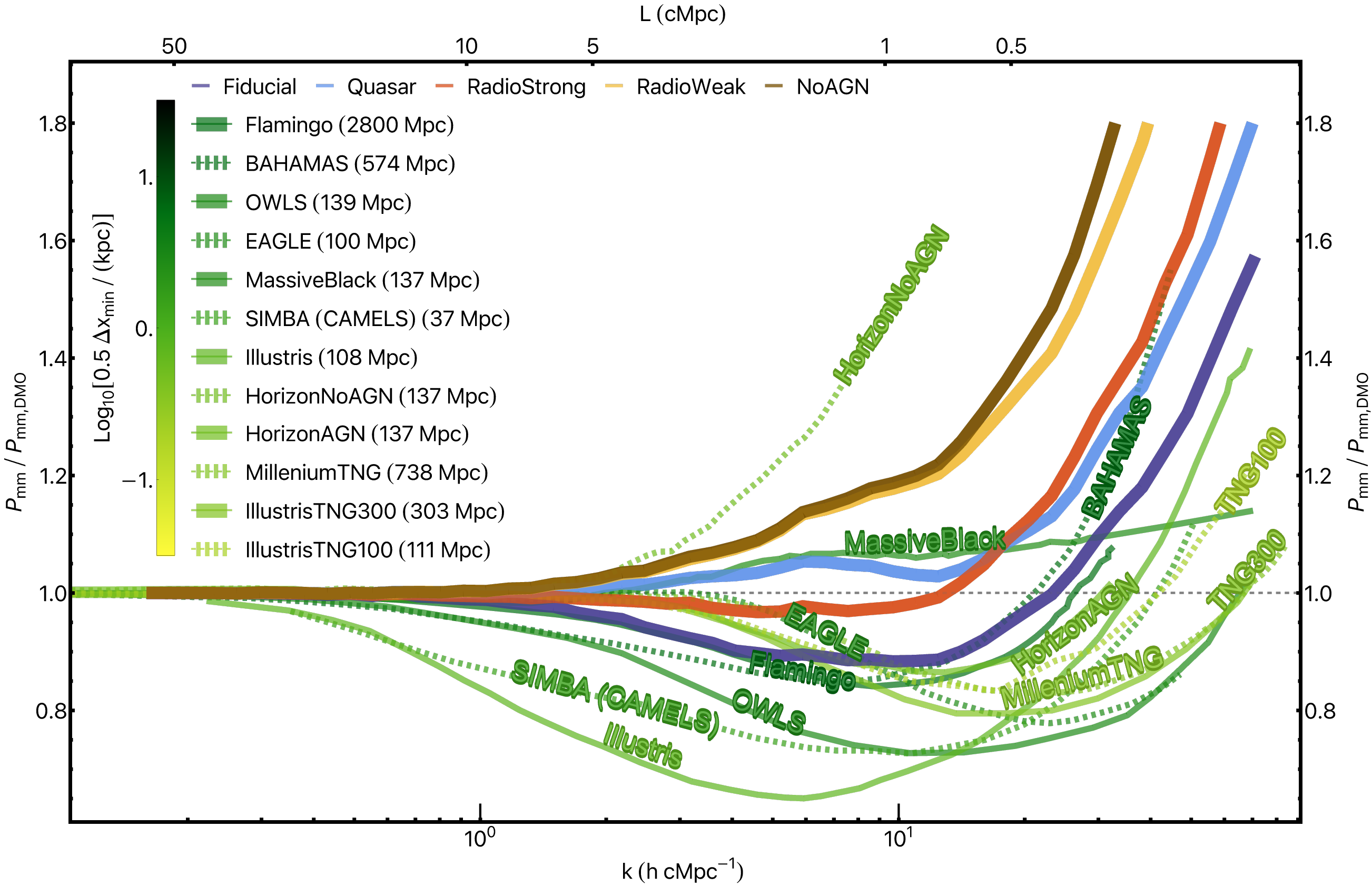}
    \caption{Fractional impact of baryons on the MPS at $z = 0$ for the main \fable~models compared with other simulations. Line colours display the minimum resolution element radius (or half-cell size, $0.5 \Delta x_\text{min}$) in each model, where darker colours correspond to higher resolution. The effects on the MPS from the \fable~\Fiducial~AGN feedback model is similar to other simulation projects, namely BAHAMAS, FLAMINGO (L2p8\_m9), IllustrisTNG-100 and HorizonAGN.}
    \label{fig:CompareMPS}
\end{figure*}

In Figure~\ref{fig:CompareMPS}, we compare the relative impact of baryons on the MPS (through $\Pmm / \PDMO$) of \fable~with their relative impact in other well-known cosmological galaxy formation simulations \footnote{Simulations originally compared by \citet{Chisari2019} are: HorizonAGN \citep{Chisari2018}, Illustris \citep{Vogelsberger2014a}, IllustrisTNG \citep{Springel2018}, OWLS \citep{VanDaalen2011}, BAHAMAS \citep{McCarthy2018}, MassiveBlack \citep{Khandai2015, Huang2019} and EAGLE \citep{Schaye2015, Hellwing2016}. We also include FLAMINGO (model L2p8\_m9; \citealt{Schaye2023}), SIMBA (as presented in CAMELS by \citealt{Villaescusa-Navarro2021}), and MilleniumTNG \citep{Pakmor2023b}}.
For each of the models, we include in parenthesis the cosmological box size, with different lines coloured according to the maximum spatial resolution of the respective simulation.
Overall, the suppression of the relative MPS in simulations due to baryonic physics, primarily due to AGN feedback, occurs at scales of $k \sim 5 - 20 \hcMpc$. Focusing on the \fable~\Fiducial~model, the scale of maximal power suppression $k_\text{peak} \sim 10 \hcMpc$ is comparable to most other simulations (typically $k_\text{peak} \sim 10 - 20 \hcMpc$) and the overall shape of the relative MPS lies within the bulk of the outcomes from other simulation projects. 

We first compare \fable~with the HorizonAGN and IllustrisTNG-100 simulations, as they have comparable order of magnitude resolutions, finding that they all have similar maximum power suppression of $\sim 10 - 20\%$.  Interestingly, \fable~has a larger impact at scales $1 < k / \hcMpc < 5$ than IllustrisTNG-100 and especially HorizonAGN, with the latter finding their relative power suppression to be concentrated at $k \gtrsim 5 \hcMpc$. Instead, AGN feedback in \fable~has a shape of the $\Pmm / \PDMO$ curve that resembles that of the OWLS or BAHAMAS simulations, but with a lower suppression magnitude. This resemblance is possibly the result of all three of these simulations employing episodic quasar mode feedback.

When comparing alternative \fable~AGN feedback physics with other simulations, the \NoFeedback~and the weakest AGN (\RadioWeak) models resemble the behaviour of the HorizonNoAGN simulation, although the \fable~cases have a less pronounced clustering at $k \gtrsim 4 \hcMpc$ probably due to stronger SN feedback. This illustrates the weak impact of \RadioWeak, despite featuring relatively unchanged galaxy populations (see \citealt{Henden2018}; or Section~\ref{ss:GalaxiesCuts}). Instead, \RadioWeak~and \NoFeedback~models have higher gas mass fractions within the largest haloes, in some tension with observations. The \Fiducial~and \RadioStrong~cases are in good agreement with observations, \citep[see Figure~A2 in][]{Henden2018}, and suggest radio mode AGN feedback in \fable~also leads to the discussed correlation between cluster gas mass fractions and MPS power suppression \citep{vanDaalen2020}. 

Another interesting comparison is that between our \Quasar~model and MassiveBlack, where both simulations show a somewhat flatter MPS, with a modest peak in clustering at $k \sim 12 \hcMpc$. MassiveBlack employs a thermal feedback prescription with a constant energy parameter of $f = 0.05$, and does not include an alternative feedback injection mechanism in the radio mode regime \citep{Khandai2015}. Consequently, it is possible that the comparable behaviour is the result of feedback being predominantly more effective at higher redshifts, whilst at low redshifts, as the BH accretion rate density decreases, the impact of AGN feedback on the MPS declines significantly. In agreement with \citet{Chisari2018}, we attribute this behaviour in \Quasar~(and possibly in MassiveBlack) to a late-time decrease in AGN regulation (due to the lack of effective 'radio-mode' feedback), where power builds up more rapidly at intermediate scales ($1 < k / \hcMpc < 10$) when gas is re-accreted into massive haloes \citep{Beckmann2017, Habouzit2021}.

Finally, we note that the \fable~AGN model has an impact well below the $\sim 40\%$ suppression with respect to the non-linear dark matter-only scenario observed for simulations such as Illustris or OWLS, which likely have ejective feedback that may be too effective \citep{Genel2014aa}.

\begin{figure*}
    \centering
    \includegraphics[width=2.1\columnwidth]{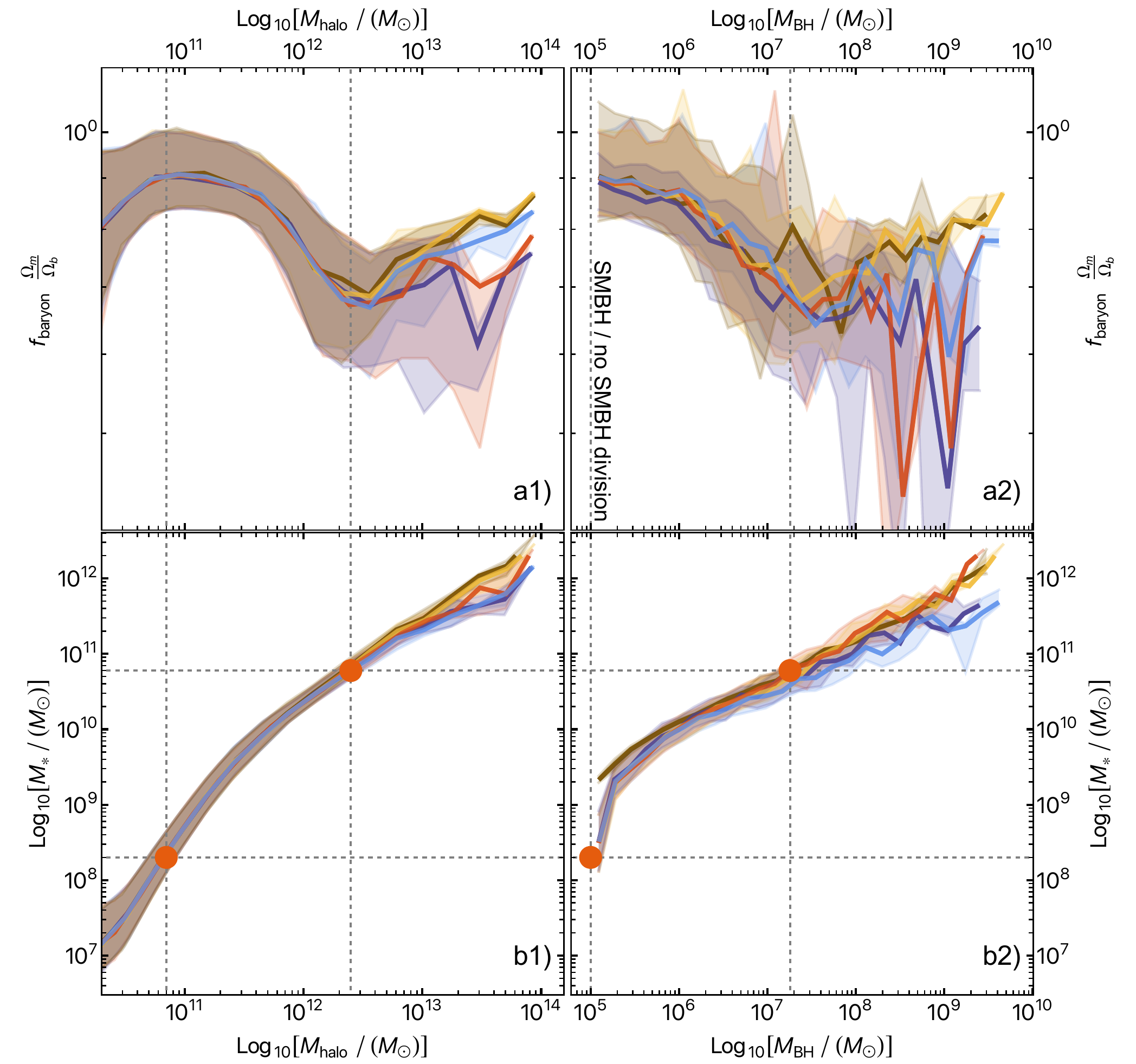}
    \rowlegendfull
    \caption{Each of the panels displays for the studied \fable~models: baryon fraction vs. halo mass {\bf (a1 panel)}, baryon fraction vs. black hole mass {\bf (a2 panel)}, 
    stellar mass - halo mass relation {\bf (b1 panel)}, and
    black hole mass - stellar mass relation {\bf (b2 panel)}. 
    Each of the solid lines indicate the median relation, with shaded bands encompassing the 20 - 80\% quantiles. 
    In all panels, vertical and horizontal dashed lines indicate the cuts employed to segregate the population of galaxies into filters for our analysis (see text). Divisions are implemented based on the virial mass and propagated to other quantities according to the population distribution. An additional cut separating galaxies with and without blackholes is used, and artificially displayed at $\sim 10^{4} \Msun$ in the b2 panel.}
    \label{fig:HaloRelations}
\end{figure*}

\begin{figure}
    \centering
    \includegraphics[width=\columnwidth]{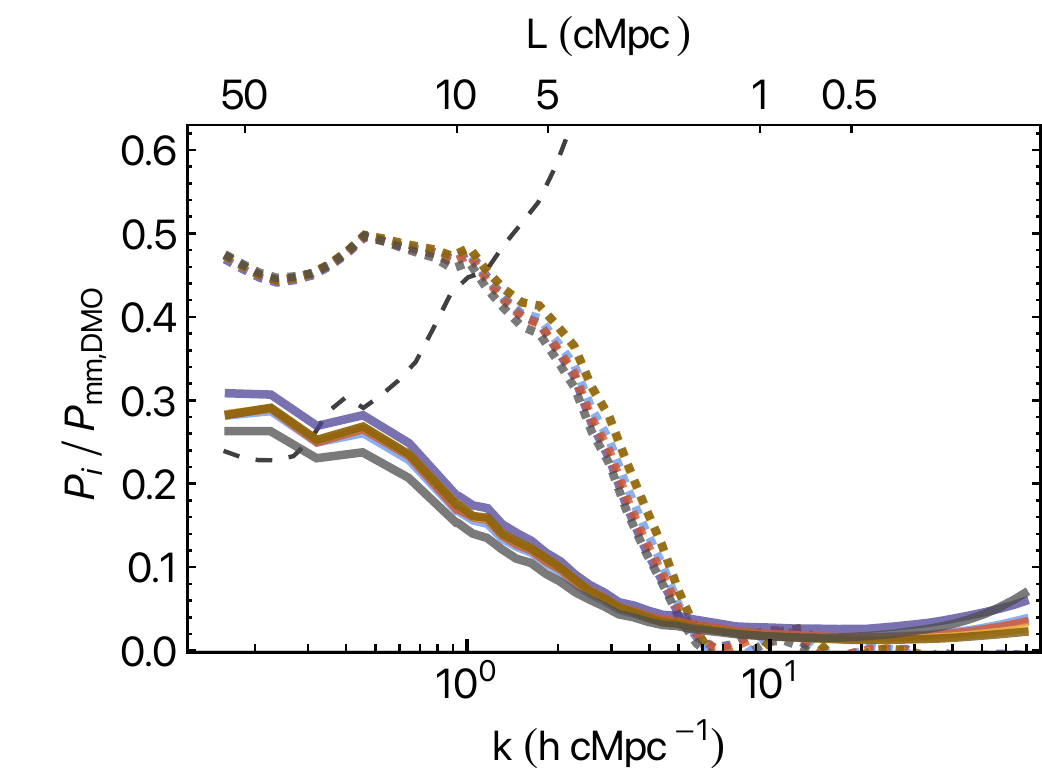}\\
    \linelegend\\
    \caption{MPS from the matter outside haloes (solid lines) and their cross correlation, $2\,\Cmfe$,  (dashed lines), divided by the total $\PDMO$. We include the same ratio for the matter {\it inside} haloes of the \DMO~simulation (thin dashed black line). This shows that the power spectra of matter within haloes dominates at small scales ($k \gtrsim 5\,\hcMpc$), whereas matter outside haloes and its cross correlation dominates power at large scales. In the \Fiducial~simulation there is a mildly higher amount of power outside haloes as this simulation model is most effective at ejecting matter from haloes.}
    \label{fig:Pmmouthaloes}
\end{figure}

\begin{figure*}
    \centering
    \includegraphics[width=2.08\columnwidth]{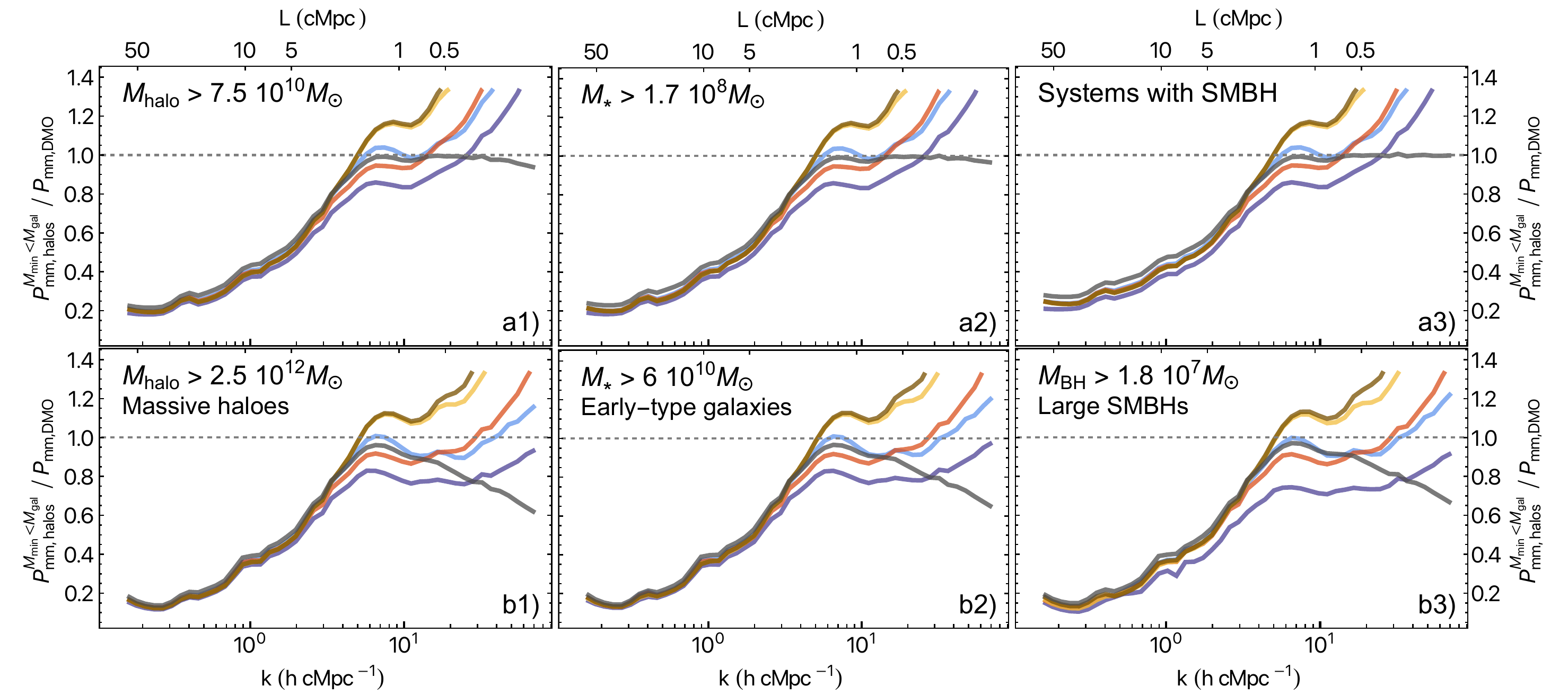}\\
    \includegraphics[width=2.08\columnwidth]{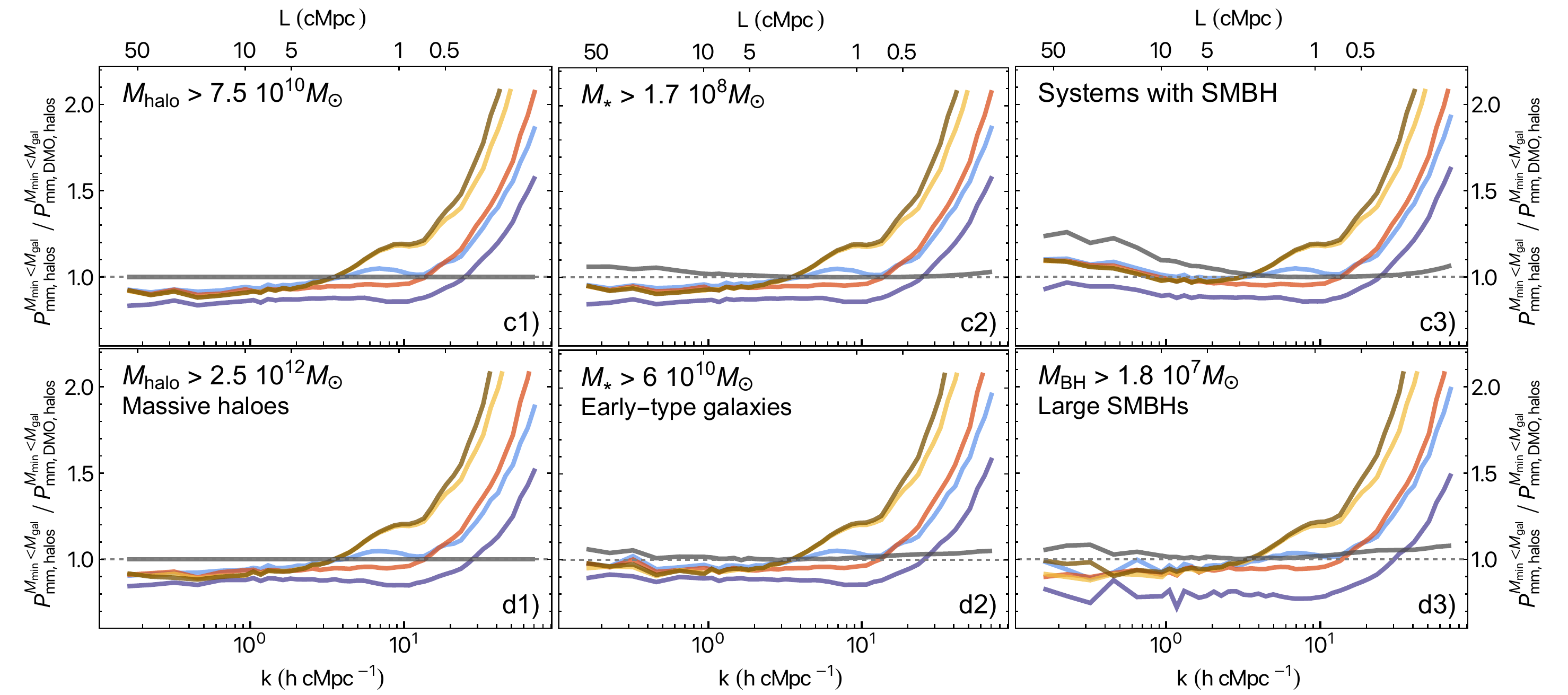}\\
    \includegraphics[width=2.08\columnwidth]{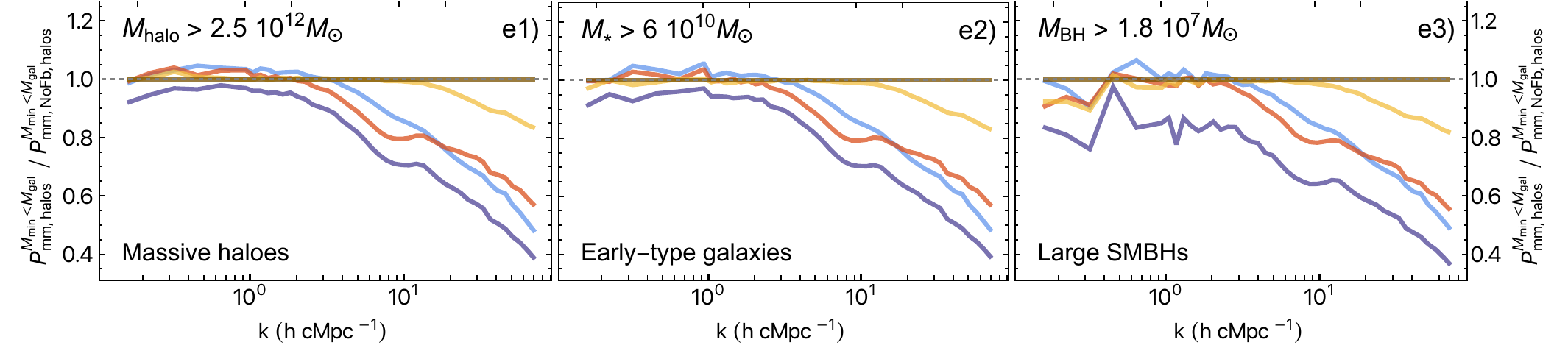}\\
    \rowlegendfull
    \caption{{\bf (Top set of panels; {\it a} and {\it b} rows)} Haloes MPS ratio with respect to the total \DMO~model MPS. From left to right, each column displays galaxy selection according to $\Mhalo$, $\Mstar$ and $\MBH$, respectively. Top and bottom rows correspond to the $\Mhalo > 7.5 \cdot 10^{10} \Msun$ and $\Mhalo > 2.5 \cdot 10^{12} \Msun$ mass thresholds, as well as their corresponding thresholds in $\Mstar$ and $\MBH$ (converted according to the galaxy distributions shown in Figure~\ref{fig:HaloRelations}).
    {\bf (Central set of panels; {\it c} and {\it d} rows)} Haloes MPS ratio as in rows {\it a} and {\it b}, but now divided by $\PDMOfilt$. Haloes in $\PDMOfilt$ are always selected according to $\Mhalo,_\text{cut}$ in all three columns.
    {\bf (Bottom set of panels; row {\it e})} Same as row {\it d}, but now divided by the $\Pmm$ of the \NoFeedback~model instead of the \DMO~simulation.
    Overall, selecting galaxies according to $\Mhalo$ or $\Mstar$ does not have a significant influence on our results, whereas a selection based on $\MBH$ leads to a higher power suppression around the hosts of the most massive black holes in \fable.
    }
    \label{fig:FilteredMPS}
\end{figure*}

\begin{figure}
    \centering
    \includegraphics[width=\columnwidth]{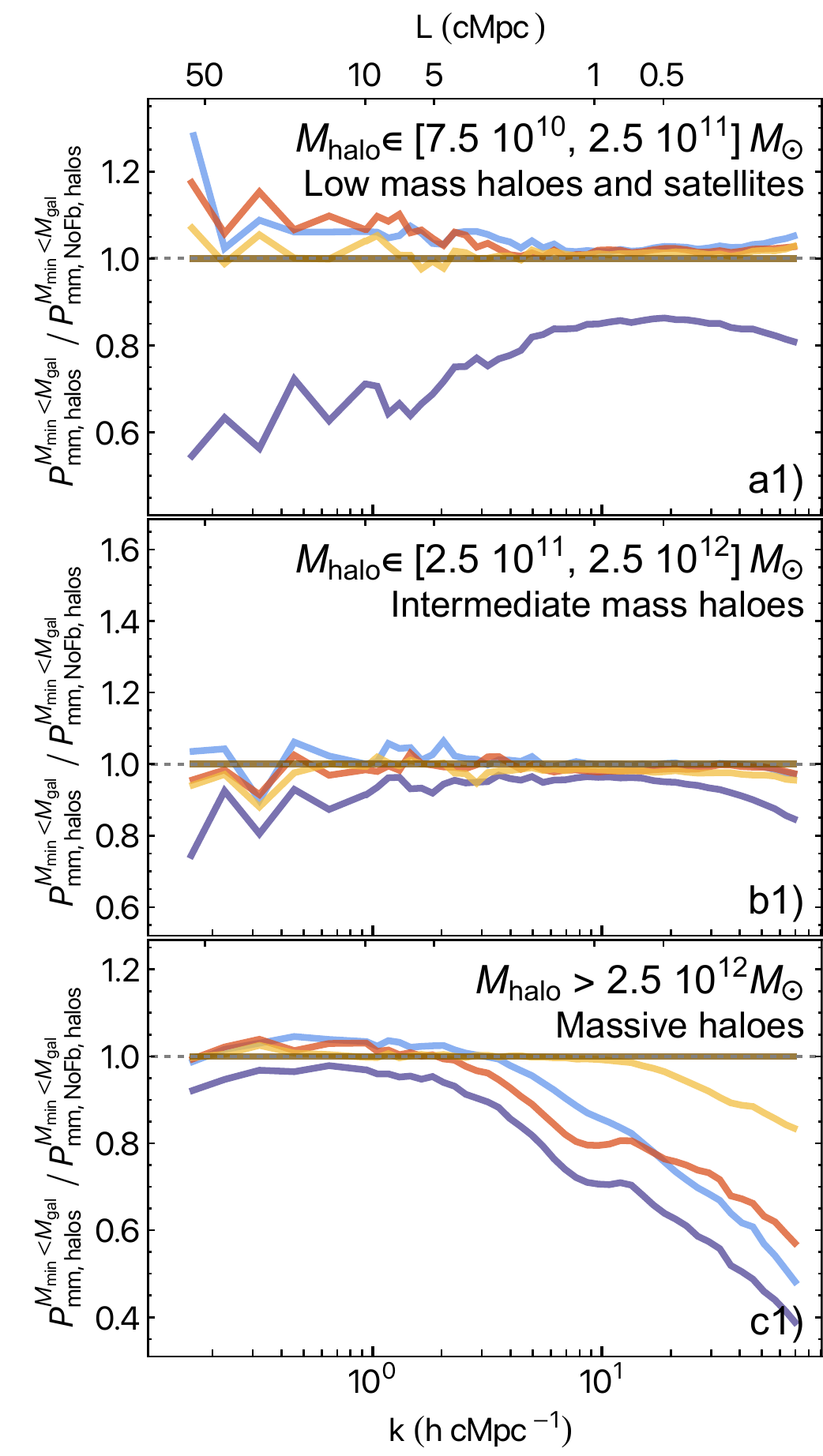}\\
    \linelegend
    \caption{Haloes MPS due to galaxy selections of low mass (top panel), intermediate mass (middle panel) and massive (bottom panel) haloes. The displayed power is divided by that of the haloes in the \NoFeedback~simulation. No significant suppression of power is observed around the smallest and intermediate mass galaxies, except for the \Fiducial~model. Due to the low mass of the $\MBH$ hosted by systems included in the top panel, we attribute the observed suppression to AGN feedback stemming from neighbouring massive galaxies.}
    \label{fig:BandsChecks_midrow}
\end{figure}

\subsection{The impact of different galaxies on the matter power spectrum}
\label{ss:GalaxiesCuts}

To understand how different galaxies and AGN feedback from their central SMBHs influences the distribution of matter in our simulations, we explore the variation of the MPS around galaxies separated according to multiple property cuts. We analyse separations according to the halo mass ($\Mhalo$), stellar mass ($\Mstar$) and black hole mass ($\MBH$) of galaxies, computed as described in Section~\ref{ss:Cuts}. The baryon fraction $\fbar$ (defined as the fraction of baryonic mass to the total mass inside each halo) is another important quantity to understand how different feedback contributes to MPS power suppression \citep[e.g.][]{Semboloni2011,Salcido2023}. We show the distribution of \fable~galaxies across the parameter space of these properties in Figure~\ref{fig:HaloRelations}. From top left to bottom right, each panel displays the $\fbar - \Mhalo$, $\fbar - \MBH$, $\Mhalo - \Mstar$ and $\MBH - \Mstar$ relations, respectively. We rescale our $\fbar$ measurements by $\Omega_m / \Omega_b$ to display the proportion with respect to the universal baryon fraction in our simulations. Solid lines in each of the panels display the median relation for each of the shown models, whereas the shaded bands encompass the 20 - 80 \% quantiles.

Overall, \fable~runs have relatively similar mean relations for the simulated galaxies, with the most important variations taking place with $\MBH$. Simulations with weaker feedback reach higher $\MBH$, but the effects of the duty cycle and increased radio mode strength affect differently systems across various $\MBH$ ranges. This effect appears more prominent on the low $\Mstar$ boundary of the $\MBH - \Mstar$ relation \citep[also see][]{Koudmani2021}. We refer the reader to \citet{Henden2018} for further analysis of the galaxy and cluster populations in \fable. When reviewing the $\fbar$ changes with respect to $\Mhalo$, we find the relative baryon content to decrease above masses of $\sim$$10^{12} \Msun$, and remain relatively flat for the stronger AGN feedback models. As expected, the relation of this same quantity with respect to the $\MBH$ reflects the AGN-driven reduction of baryonic content in these haloes, as we find a clear decrease of $\fbar$ with increasing $\MBH$ for increasingly stronger AGN feedback models.

Figure~\ref{fig:HaloRelations} also includes dashed lines corresponding to the cuts employed to separate our galaxies into different mass ranges. We present only a subset of all the investigated cuts, varying these cuts results only in monotonic and minor variations. Our mass range divisions are first set to separate halo masses (i.e., $\Mhalo,_\text{cut} \in \left[7.5 \cdot 10^{10}, 2.5 \cdot 10^{11}, 7.5 \cdot 10^{11}, 2.5 \cdot 10^{12}\right] \Msun$). These $\Mhalo,_\text{cut}$ are then converted to $\Mstar,_\text{cut}$ and $\MBH,_\text{cut}$ following the population medians of the \Fiducial~\fable~scalings. The intersects across such scalings are shown as orange points in Figure~\ref{fig:HaloRelations}, and the resulting range division values correspond to $\Mstar,_\text{cut} = \left[1.7 \cdot 10^8, 2.5\cdot 10^9, 1.5\cdot 10^{10}, 6.0\cdot 10^{10}\right] \Msun$ and $\MBH,_\text{cut} = \left[0.0,\, 2.5\cdot 10^5, 1.4\cdot 10^6, 1.8\cdot 10^7\right] \Msun$, where the first cut in $\MBH$ separates galaxies with and without a SMBH. These divisions serve to review comparable populations selected under different criteria.

To illustrate the effects of filtering our mass distribution, Figure~\ref{fig:Pmmouthaloes} presents a comparative analysis of the MPS inside and outside of haloes ($\Mhalo > 7.5 \cdot 10^{10} \Msun$), along with twice their cross-correlation (i.e. $2\,\Cmfe = \Pmm - (\Pmf + \Pme)$). The volume outside of haloes constitutes most of the simulation domain, and altogether with the cross correlation, dominates the contribution to $\Pmm$ at $k < 2\hcMpc$. At these large scales, haloes contribution to $\Pmf / \PDMO$ is of order of $\sim 0.2$. 

 Across different AGN models, variations in $\Pme$ and $\Cmfe$ are relatively minor, with only a slightly higher amount of power outside of haloes in \Fiducial~and \RadioStrong. Models with stronger radio mode AGN feedback also display somewhat lower cross-correlation in the $10 \lesssim k / (\hcMpc) \lesssim 4$ range, with this quantity rapidly approaching zero towards smaller-scales
 . It will be important for the analysis below to emphasise that at scales $k \gtrsim 7\,\hcMpc$, the mass outside of haloes and its cross-correlation with $\Pmf$ only constitutes a small fraction of $\Pmm$ ($\lesssim 0.1 \PDMO$). Furthermore, differences across models are smaller than the variations observed in $\Pmf$ or in Figure~\ref{fig:CompareMPS}. Consequently, any suppression observed in $\Pmm$ emerges from variations of the power within haloes.

We now focus on the variations of the MPS within haloes, selected according the thresholds describe above. Figure~\ref{fig:FilteredMPS} shows the resulting spectra $\Pmf$, for our lowest and highest threshold selections: $\Mhalo > 7.5 \cdot 10^{10}~\Msun$ and $\Mhalo > 2.5 \cdot 10^{12}~\Msun$ (leftmost column), as well as their corresponding thresholds in $\Mstar$ and $\MBH$. From left to right, columns correspond to cuts on $\Mhalo$, $\Mstar$ and $\MBH$. The top set of panels show the filtered spectra divided by the total MPS of the \DMO~model ($\Pmf / \PDMO$). For a better view of variations across \fable~models, the central set of panels shows the same mass thresholds, now divided with respect to the halo mass filtered \DMO~simulation ($\Pmf / \PDMOfilt$). Consequently, the ratio of all panels in a row is computed with respect to the $\PDMOfilt$~in its leftmost column. This allows for a direct comparison across selection masses and \fable~models. Finally, and to facilitate further comparison with Figure~\ref{fig:PkRatioDM}, we show in the bottom row the ratio of $\Pmf$ with respect to the $\Pmf$ of the \NoFeedback~simulation.

The differences across the various models, cuts, and selection mass types are more prominent at small scales ($k > 5\,\hcMpc$). When all galaxies are considered ({\it a} row panels) a trend towards higher clustering of the hydrodynamical runs with respect to the haloes in the \DMO~model is observed. Such transition from lower to higher clustering occurs at different scales for different models, depending on the efficiency of AGN feedback ({\it a1} and {\it c1} panels). For \NoFeedback, \RadioWeak, and \Quasar, this occurs at $k \sim 4 \, \hcMpc$. For \RadioStrong~and \Fiducial~it takes place at smaller scales, with $k \sim 12 \, \hcMpc$ 
and $\sim 25 \, \hcMpc$, respectively. As haloes are progressively discarded by increasing the threshold mass, the amount of small-scale power is progressively reduced until only the largest haloes are considered (as shown in {\it b} row panels). Comparing the least and most restrictive threshold, we find a considerable decrease ($\sim 0.3 \PDMO$) of power at scales $k \gtrsim 20 \hcMpc$, but a negligible reduction in the $k \in [5 - 10]\,\hcMpc$ range. The proportional separation between the $\Pmf$~of the hydrodynamical models and $\PDMOfilt$~varies differently when increasing mass cuts ({\it c} vs {\it d} rows), with AGN feedback affecting differently the power clustering contribution from different halo masses \citep{vanLoon2023}. For example, $\Pmf$ of \Fiducial~is reduced from about 20\% to 10\% over $\PDMOfilt$ at scales of $k \sim 30\, \hcMpc$, whereas \RadioWeak~increases from 55\% over to 65\% over $\PDMOfilt$. We attribute this to AGN feedback in this model becoming less capable of impacting more massive galaxies (panels {\it c1} vs {\it d1}). Despite their differences, the scale at which these weaker AGN models (i.e., \NoFeedback, \RadioWeak, and \Quasar) transition from having less power than the \DMO~case ($\Pmf / \PDMOfilt < 1$; large scales) to more power than the \DMO~case ($\Pmf / \PDMOfilt > 1$; small scales) remains approximately unchanged at a scale of $k \sim 4 \, \hcMpc$. This scale is approximately independent of the mass cut employed. For the models with strongest AGN feedback (\RadioStrong~and \Fiducial), this scale remains unchanged for $\Mhalo$ and $\Mstar$ sample selections, but shifts to almost a factor $2\times$ higher scales when the galaxies hosting the most massive black holes are selected (panels {\it c3} vs {\it d3}). Applying the most stringent mass cut according to $\MBH$ rather than $\Mhalo$ (or $\Mstar$) enhances differences between models. This $\MBH$-based selection further increases the power suppression observed due to AGN across all scales (panel {\it d3}). This is also in agreement with the expectation of higher power suppression with decreasing $\fbar$, as selecting galaxies in our \Fiducial~model with increasingly larger $\MBH$ leads to a faster decrease of $\fbar$ than when the analogous selection is applied on $\Mhalo$ (Figure~\ref{fig:HaloRelations}, panels a1 and a2).

Such lack of variations depending on whether $\Mhalo$ and $\Mstar$ sample selection criteria are employed reflects a tighter interrelation between these two quantitites at the high mass end of the galaxy population, which dominate the $\Pmf$. The enhanced relative power suppression when massive SMBH are selected confirms the mass of SMBHs as an integrated measure of AGN feedback and power suppression. However, the specific implementation of AGN feedback may significantly affect the amount of the MPS suppression, motivating further exploration of more sophisticated AGN model implementations \citep[e.g.,][]{Zubovas2016, Bourne2017, Costa2020, Talbot2021, Beckmann2022a, Husko2022, Koudmani2023, Rennehan2023}.

When reviewing such differences across models, the trends observed in Figure~\ref{fig:CompareMPS} are clearly reflected in the $\Pmf / \PDMO$ ratio at scales $k \gtrsim 7\,\hcMpc$. This is true both for the ratio with respect to $\PDMO$ ({\it a} and {\it b} rows) and with respect to the \NoFeedback~model ({\it e} row). As discussed above (Figure~\ref{fig:Pmmouthaloes}), haloes dominate MPS power at such scales. Consequently, the suppression observed in $\Pmm$ emerges necessarily from the effects of AGN inside haloes. The lower power across all scales in \Fiducial~({\it c} row panels) and \RadioStrong~to a lower extent, is the result of clustering being reduced at small scales and matter being ejected outside of haloes \citep{vanLoon2023}. This is in agreement with the slight increase of power outside of haloes (Figure~\ref{fig:Pmmouthaloes}; top panel) at scales of $k \sim 10\,\hcMpc$. As above, whether systems are selected according to their $\Mstar$ or their $\Mhalo$ does not have a significant effect on these results. Selecting systems according to the SMBH mass accentuates differences between models, with the \Fiducial, \RadioStrong, and \Quasar~models exhibiting a lower amount of power at scales $k \gtrsim 2\,\hcMpc$. The resulting selection of systems preferentially focuses on haloes from which a larger proportion of matter has been removed or ejected \footnote{We confirmed this is not the result of a lower number of systems being included when performing our analysis assuming the SMBH threshold, where most models actually include slightly more galaxies.}. Finally, the peak of power at $k \sim 6 - 7\,\hcMpc$ in the \Quasar~model is somewhat suppressed by the SMBH selection. We attribute the peak to more inefficient feedback around less massive SMBHs: as the \Quasar~simulation transitions from a more prevalent quasar mode at high redshift to radio mode dominance (Figure~\ref{fig:AGNmodeRedshift}), clustering increases around massive haloes with AGN less efficient in mass removal \citep{Beckmann2017}.

To provide further insight into the effect of feedback around smaller galaxies, we repeat a similar analysis now selecting only galaxies within non-intersecting mass intervals, instead of a minimum mass threshold. This isolates their power contribution, as they are secondary when compared with the most massive systems \citep{vanDaalen2015}. The resulting spectra for different haloes are shown in Figure~\ref{fig:BandsChecks_midrow}, employing a $\Mhalo$ selection. The spectra of haloes is shown as the ratio of each model with respect to the \NoFeedback~case. Most models do not have any significant effect on $\Pmf$ around the smallest galaxies (top panel) except \Fiducial. In this simulation, a large suppression takes place across all studied scales, and especially at $k \lesssim 1\, \hcMpc$. Due to these systems being hosts of small $\MBH$, with low integrated power budgets, the observed power suppression is potentially driven by AGN residing in large neighbouring galaxies affecting their satellites and nearby smaller galaxies \citep{Dashyan2019,Martin-Navarro2021,Shuntov2022}. The lack of any significant suppression around intermediate mass haloes ($2.5 \cdot 10^{11}~\Msun < \Mhalo < 2.5 \cdot 10^{12}~\Msun$; central panel) supports a scenario where only the largest SMBH are capable of such clustering suppression. Once again, only the \Fiducial~simulation experiences some notable clustering reduction, notable at the smallest scales. This lack of suppression around haloes $\Mhalo < 5 \times 10^{12}\Msun$ was also found by \citet{vanLoon2023}. This behaviour remains approximately unchanged regardless of whether an $\Mhalo$, $\Mstar$, or $\MBH$ selection is employed.

\begin{figure}
    \centering
    \includegraphics[width=0.9\columnwidth]{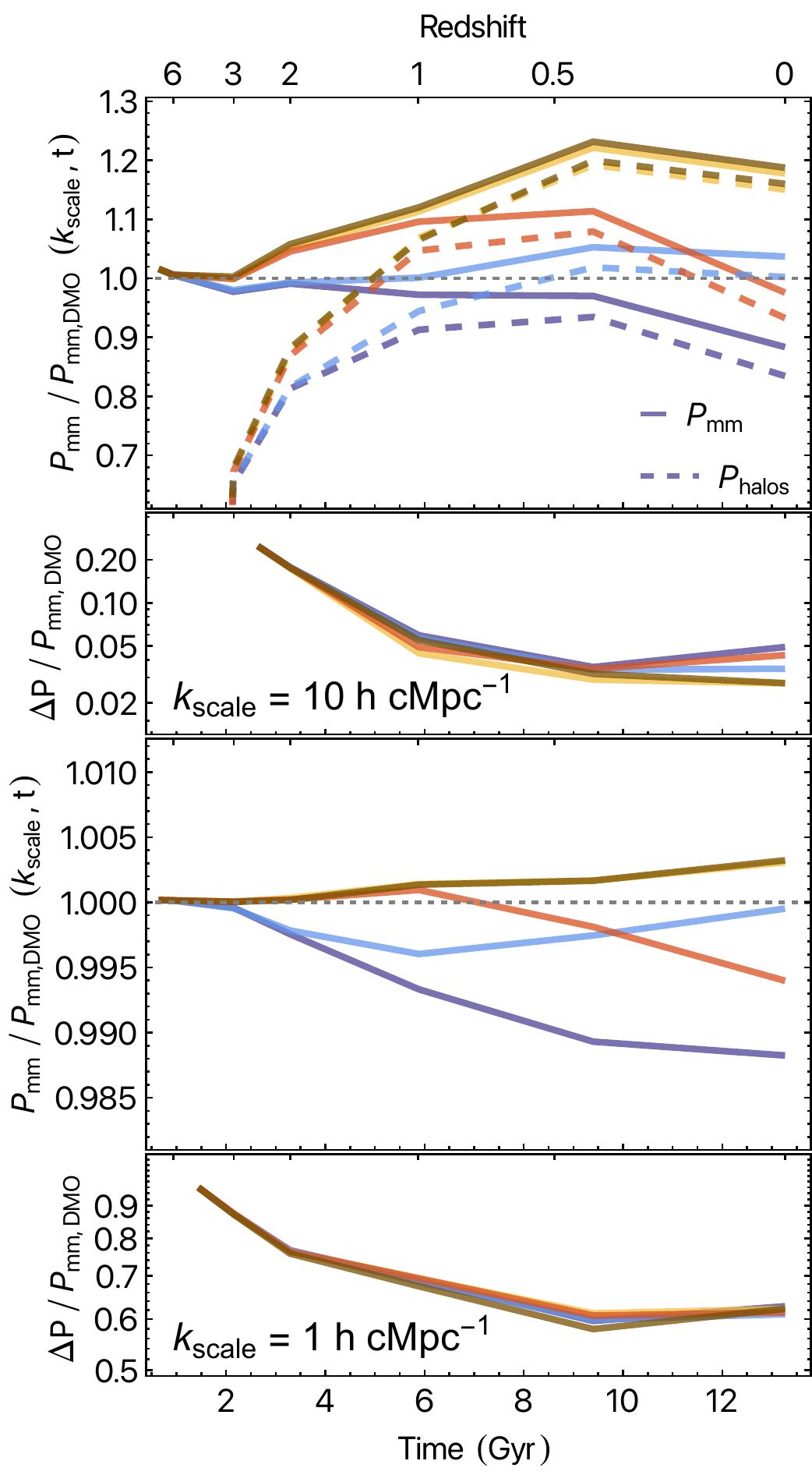}
    \linelegend
    \caption{{\bf (First and third panels)} Time evolution of the fractional impact of baryonic physics on the MPS at scales of $\kscaleten$ and $\kscaleone$, respectively. The solid lines displays $\Pmm$, whereas the dashed line correspond to the MPS of all matter within all haloes $\Pmmhalo$. Note that the $\Pmmhalo$ is not shown in the third panel, as it falls well below the displayed range. {\bf (Second and fourth panels)} relative difference between $\Pmm$ and $\Pmmhalo$ ($\Delta P = \Pmm - \Pmmhalo$) for each model, at both $\kscaleten$ and $\kscaleone$, respectively. The impact of AGN feedback on $\Pmmhalo$ is responsible for most suppression of power observed in $\Pmm$ for $\kscaleten$ at $z \lesssim 1$. At higher redshifts, the contribution of baryonic effects outside haloes becomes important ($>10\%$). Precision modelling of $\Pmm \left(\kscaleone\right)$ in the \fable~suite also requires accurate characterisation of AGN feedback effects both within and outside haloes.}
    \label{fig:HalosTimeSuppression}
\end{figure}

\subsection{Tracing AGN power suppression with haloes at different scales and times}
\label{ss:HaloesTrace}

To further understand how baryonic physics modifies clustering inside and outside of haloes across cosmic time, we show the evolution of $\Pmm (k_\text{scale}) / \PDMO$ in Figure~\ref{fig:HalosTimeSuppression}. The top panel displays our results at $\kscaleten$. At early times ($z \gtrsim 1$), the power inside of haloes is the primary contribution to the total $\Pmm$, but is not fully dominant. The \Quasar~and \RadioStrong~models reveal how clustering is sensitive to the efficiency of the radio versus quasar modes, both inside and outside haloes. The second panel shows the ratio of the difference $\Delta P = \Pmm - \Pmmhalo$, which highlights how well is the total MPS traced by $\Pmmhalo$ for each model. By $z \sim 1$, haloes constitute approximately 95\% of $\Pmm$, and despite significant variations in total power within haloes across models, the proportion of power outside of haloes remains similar across simulations. 

Hence, in our hydrodynamical simulations, any large deviations of $\Pmm$ (or $P_\text{mm,haloes}$) from $\Pmm$ in the \NoFeedback~simulation is primarily driven by AGN feedback modifying the clustering of matter. With the \Quasar~and \Fiducial~models evolving comparably, the lower $\Pmm$ at $z \sim 0$ due to the higher efficiency of the implemented quasar duty cycle is driven by a suppression of power inside haloes. In the \Quasar~model, the enhanced impact due to the quasar duty cycle at early times preserves its imprint down to $z = 0$. After $z \sim 0.5$, the enhanced radio feedback power in the \Fiducial~and \RadioStrong~simulations leads to a considerable suppression of halo clustering. This drives a late-time decrease of $\Pmm$ in both models. At $z \lesssim 1$, the difference between the total $\Pmm$ and that of all haloes remains considerably smaller than the deviations across models, and further confirms the impact of baryons at $k_\text{scale} \gtrsim 10 \hcMpc$ to be driven by AGN feedback, and constrained to the interior of haloes.

At large scales ($\kscaleone$; third and fourth panel), and subject to the caveats described in Section~\ref{s:Caveats}, redshift evolution of all models resembles their $\kscaleten$ counterparts. \Quasar~and \Fiducial~feature some power suppression until $z \sim 1$. However, the effect of the enhanced radio mode is already in place by $z \sim 1$ (instead of $z \sim 0.5$ for $\kscaleten$). It leads to a suppression of power in \Fiducial~only slightly above 1\%. As shown in Figure~\ref{fig:Pmmouthaloes}, and reflected by the bottom panel, power at this scales is dominated by matter outside of haloes and its cross-correlation with galaxies. With variations in the fraction of power outside of haloes (bottom panel) being larger than the actual suppression, understanding the power decrease with respect to the \DMO~requires constraining power both inside and outside of haloes. Overall, Figure~\ref{fig:HalosTimeSuppression} illustrates how different AGN implementations will not only modify the details of $\Pmm$ and $\Pmmhalo$ at $z = 0$, but also its redshift evolution. Hence, any precision measurements ($< 1\%$) will require a more detailed understanding of how AGN feedback operates across cosmic time \citep{Semboloni2011, Huang2019, Chisari2019}. For the interested reader, the time evolution of $\Pmmhalo$ across all studied scales is shown in Appendix~\ref{appsec:HalosEvolution}.

\begin{figure}
    \centering
    \includegraphics[width=\columnwidth]{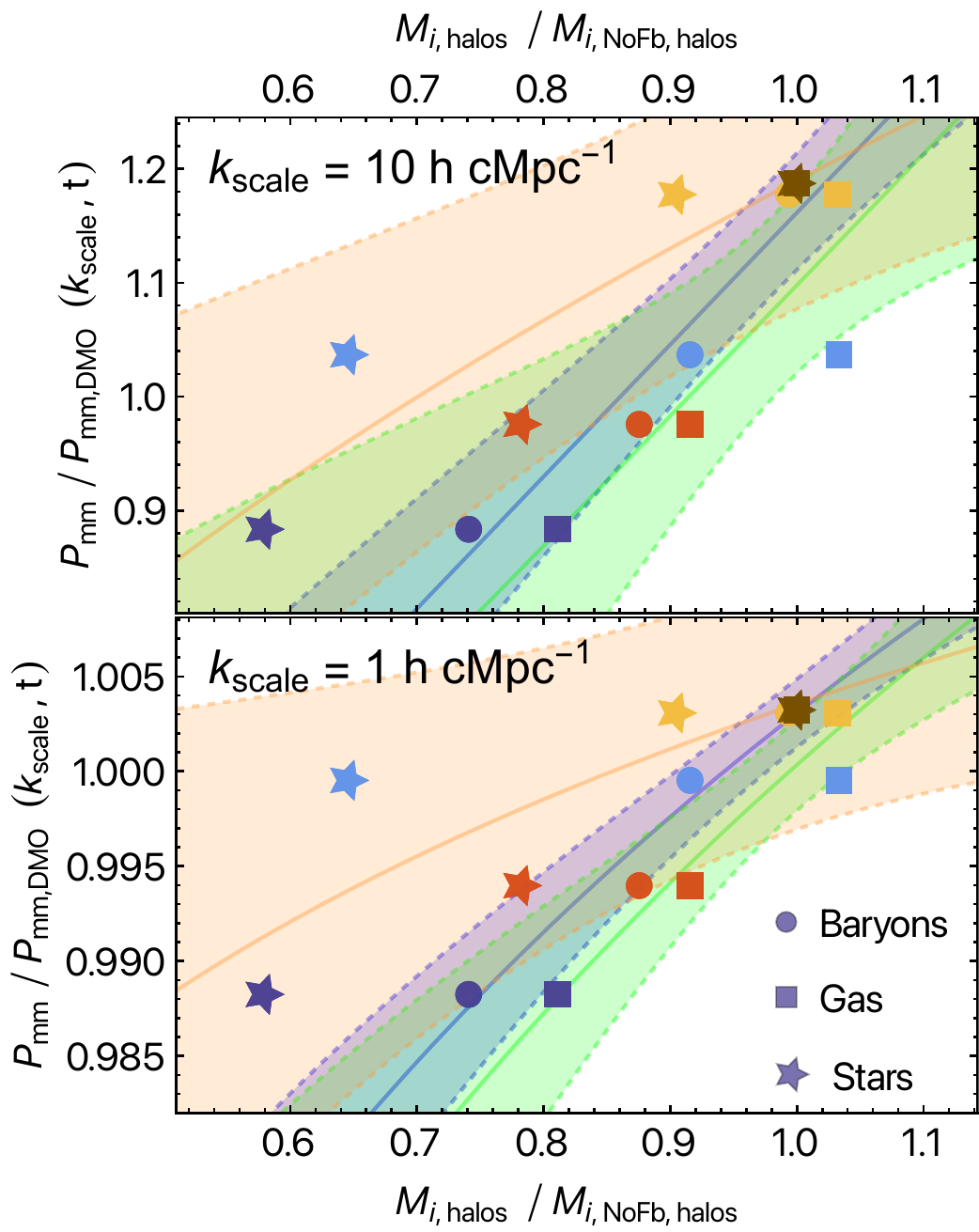}\\
    \includegraphics[width=\columnwidth]{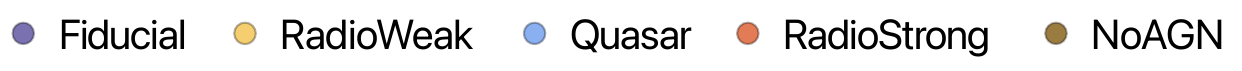}\\
    \vspace{-0.2cm}
    \caption{Fractional impact of baryonic physics for the different AGN models as a function of the relative mass within haloes with respect to the \NoFeedback~simulation. Different symbols represent this quantity for the stellar mass (star symbols), gas mass (square symbols), and total baryonic mass (including SMBH mass; circle symbols). Different \fable~runs are displayed with different symbol colours  (see legend below panels). We also include separate power-law best fits (see text) to the stars, gas and total baryons symbols, as yellow, green and blue bands, respectively.  Relative baryonic mass within haloes provides the best tracer for the relative suppression of power, both at $\kscaleten$ and $\kscaleone$.}
    \label{fig:HalosContentSuppression}
\end{figure}

We conclude by revisiting in Figure~\ref{fig:HalosContentSuppression} how the baryon content in the most massive galaxies, as well as that of the gas and stellar mass, correlates with power suppression at each of these two scales in \fable~simulations. These are selected as $\Mstar > 6 \cdot 10^{10} \Msun$, to investigate how well the most massive systems trace the power suppression of our different AGN models. Their power suppression is shown as a function of the total mass in the haloes of all selected systems ($M_{i, \text{halos}}$), separately for baryons (circles), gas (squares) and stars (star symbol). We normalise $M_{i, \text{halos}}$ to their values in the \NoFeedback~simulation ($M_{i, \text{halos}} / M_{\text{NoFb}, \text{halos}}$). In addition, we calculate power-law best fits to the stellar measurements (yellow band), gas measurements (green band) and total baryonic mass measurement (blue band). Note that due to too high star formation rates in massive haloes in \NoFeedback~model, its gas content is depleted and is somewhat lower than that for \RadioWeak~and \Quasar~simulations.
The total baryonic component provides an accurate tracer of power suppression at both $\kscaleten$ and $\kscaleone$ (coefficient of determination $R^2 \sim 0.95$), in agreement with previous work  \citep[e.g.,][]{Semboloni2011, McCarthy2018, vanDaalen2020, Debackere2020, Salcido2023}. Separate baryonic mass components provide a less tight constraint on the amount of suppression explored at these two different scales. Focusing on the specific baryonic components, the gas mass provides a better tracer at $\kscaleone$ ($R^2_\text{gas,1} \sim 0.89$; $R^2_\text{stars,1} \sim 0.62$), whereas the stellar mass performs better at scales of $k \sim 10\,\hcMpc$ ($R^2_\text{gas,10} \sim 0.73$; $R^2_\text{stars,10} \sim 0.77$).

Comparing now the different AGN models, note that the quasar mode duty cycle provides a mechanism that efficiently suppresses the stellar mass of massive haloes with a lower gas ejection than the enhanced radio mode. Hence, the \Quasar~simulation has a higher suppression at small scales $k_\text{scale} \sim 10\,\hcMpc$ (closer to \RadioStrong), whereas it is closer to the \RadioWeak~case at large scales ($k_\text{scale} \sim 1\,\hcMpc$). As a result, its large suppression of stellar mass provides a better correlation at $\kscaleten$, where it displays more clustering suppression. \RadioStrong~and \Quasar~feedback models displace the correlation in different directions, once again illustrating how different AGN implementations may allow to modify galaxy properties, mass content and the impact of baryons on the $\Pmm$ separately. 

\section{Caveats and considerations}
\label{s:Caveats}

While the studied \fable~simulations provide a high-resolution  model for AGN feedback in galaxies and clusters, and their backreaction on the MPS, the comparatively small box size (i.e., 40~cMpc~$h^{-1}$) leads to a number of important shortcomings that should be considered.

As a result of simulation boxes having a limited sampling of their largest wavelength modes, domains with modest cosmological sampling feature a low representation of relevant domain modes, which leads to an underestimation of the MPS power at large scales \citep{vanDaalen2015, Schaye2023}. This shortcoming may be important down to scales of $k_\text{peak} \lesssim 0.02\,\hcMpc$, with MPS power only decreasing at $k < k_\text{peak}$.
For our simulated domain, this limited sampling of large modes with wavenumber below a few~$\hcMpc$ leads to an underrepresentation of massive haloes, as well as the proportion of the total mass that is contained within this systems. For the specific box size of the \fable~simulations studied here, this corresponds to galaxy clusters with halo masses of $\gtrsim 10^{13} \Msun$ \citep{vanDaalen2015}. Importantly, at scales $k \sim 1 - 5\,\hcMpc$, clusters with such masses contribute most of the MPS power 
\citep[see e.g.,][]{vanLoon2023}.
Previous studies identify such massive clusters, specifically those with $\Mhalo \sim 10^{14}\,\Msun$, to dominate the suppression of power in the MPS \citep[e.g.,][]{vanDaalen2020, Salcido2023}.  
These caveats emerging from the relatively small box size of our models imply that our results in the $k \lesssim 1 \, \hcMpc$ regime should be interpreted with particular care. 

Despite these caveats, we note that the primary goal of this investigation is highlighting how variations of AGN feedback modelling in the high resolution regime leads to important changes in the baryonic feedback of galaxies and clusters on the MPS. A broad comparison of our fiducial model with larger simulations further places this variability into context, showing that it follows consistent qualitative trends to models such as BAHAMAS or FLAMINGO; while featuring a higher power suppression at scales of $k \sim 1\,\hcMpc$ than other models such as IllustrisTNG300 or MilleniumTNG.
Finally, \citet{Bigwood2025} find the \fable~\Fiducial~model has a comparable impact on the MPS across different cosmological seeds and simulated domain sizes spanning from the studied 40 cMpc $h^{-1}$  up to 100 cMpc $h^{-1}$, particularly at scales $k \lesssim 1 \,\hcMpc$, further supporting the qualitative robustness of our conclusions down to scales of $\sim 1\, \hcMpc$. \citet{Bigwood2025} also reveal that at scales of $k \sim 10 \,\hcMpc$, cosmic variance can lead to variations of relative power suppression between $0.86 - 0.91$, with the \Fiducial~\fable~model studied here being representative of the spread and average of their new simulations.

\section{Conclusions}
\label{s:Conclusions}
In this work we study how variations in the radio and quasar mode around the fiducial \fable~AGN model \citep{Henden2018} impact the distribution of matter at different cosmic times. The \fable~simulations are performed with the {\sc arepo} code \citep{Springel2010aa}, evolving a uniform cosmological box with $40\, \text{h}^{-1}\, \text{Mpc}$ on a side and featuring a galaxy formation model following {\sc Illustris} \citep{Vogelsberger2014b, Genel2014aa, Sijacki2015}. In addition to a dark matter-only simulation, the studied suite of 5 models spans: no AGN feedback (\NoFeedback), weak AGN feedback (\RadioWeak), stronger AGN radio mode (\RadioStrong), a quasar mode duty cycle (\Quasar), and a fiducial model combining the stronger radio mode with the quasar duty cycle (\Fiducial).

For each of these models, we investigate the matter power spectrum (MPS) and how different haloes selected accordingly to varying $\Mhalo$, $\Mstar$ and $\MBH$ thresholds contribute to it. Our main findings are summarised as follows:
\begin{itemize}
    \item The \fable~AGN models feature the largest MPS power suppression at scales of $k \sim 10\,\hcMpc$ and at $z = 0$, with a reduction of $\sim 10\%$ with respect to the \DMO~scenario. At $k \sim 1\,\hcMpc$, the \Fiducial~model has a clustering suppression of $\sim 0.012 \PDMO$. The impact of baryonic feedback on the MPS in the \fable~\Fiducial~simulation is in general comparable to Horizon-AGN and IllustrisTNG-100, but is more similar to the BAHAMAS and Flamingo simulations at scales $k \lesssim 5\,\hcMpc$. 
    
    \item Stronger radio mode feedback (\RadioStrong) is more effective at suppressing power at large scales (particularly $k \lesssim 5 \hcMpc$) and at late cosmic times ($z \lesssim 1$). The effects of the quasar duty cycle (\Quasar) are complementary to this, being more effective at smaller scales ($k \gtrsim 10 \hcMpc$), and with their most important impact at early cosmic times ($3 < z < 1$). Variations in these two modes allow for comparable impacts at $z \lesssim 0.5$, but importantly lead to significantly different redshift evolution up to $z \sim 3$, which future observations probing into the high-redshift regime will be able to constrain \citep{Huang2019}.
    
    \item Clustering suppression in \fable~takes place around the most massive galaxies ($\Mhalo > 2.5 \times 10^{12}\Msun$). This is approximately unchanged whether galaxies are selected employing halo or stellar masses. Smaller galaxies display no significant MPS suppression, except for small haloes ($\Mhalo \sim 10^{11}\Msun$) in the \Fiducial~simulation, which are likely satellites or neighbours of massive galaxies hosting large SMBH. Interestingly, selecting haloes above a given central SMBH mass threshold leads to the highest amount of relative MPS power suppression with respect to the \DMO~scenario, particularly at scales $k \gtrsim 7 \hcMpc$. 

    \item The baryonic impact on the MPS at scales of $k \gtrsim 10\,\hcMpc$ is primarily due to clustering suppression within haloes at $z \lesssim 1$. At higher redshift, and larger scales, power suppression comes from a combination of modifications of the matter distribution both inside and outside of haloes.

    \item The total baryonic mass content in the most massive haloes of the \fable~simulation provides an accurate tracer of MPS power suppression both at large ($\kscaleone$) and small ($\kscaleten$) scales \citep{vanDaalen2020}. However, modifications in the quasar and radio AGN modes drive the correlation of the stellar and gaseous components in opposite directions.  
\end{itemize}   

Overall, our results illustrate how different AGN implementations, and especially variations of the quasar and radio mode feedback, have distinct effects on the distribution of matter and the MPS redshift evolution. 
Our work motivates further delving into resolving the AGN feedback physics occurring on small scales ($\lesssim 1$~kpc) and capturing it within larger computational domains to better understand how these results vary in more representative cosmological domains (\citealt{vanDaalen2015, Salcido2023}; see Section~\ref{s:Caveats}). Additionally, it also supports
further exploration of more sophisticated and physically-motivated feedback models, either through enhanced resolution in the innermost regions of galaxies \citep[e.g.][]{Curtis2016, 
BourneEtAl2019, Beckmann2019, BourneSijacki2021, Martin-Alvarez2022, Hopkins2024}, more realistic modelling of SMBH accretion and AGN activity \citep[e.g.][]{Bourne2017, Talbot2021, Husko2022, Koudmani2023, Rennehan2023}, or even through the inclusion of non-thermal components \citep[e.g.][]{Pfrommer2017b, Costa2018b, Martin-Alvarez2021, Su2021, Beckmann2022a, Wellons2023, Ruszkowski2023}.

\section*{Acknowledgements}
We kindly thank the referee for their insightful comments and suggestions that helped improve the quality of this work.
S.M.A. is supported by a Kavli Institute for Particle Astrophysics and Cosmology (KIPAC) Fellowship, and by the NASA/DLR Stratospheric Observatory for Infrared Astronomy (SOFIA) under the 08\_0012 Program. SOFIA is jointly operated by the Universities Space Research Association, Inc. (USRA), under NASA contract NNA17BF53C, and the Deutsches SOFIA Institut (DSI) under DLR contract 50OK0901 to the University of Stuttgart. S.M.A acknowledges visitor support from the Kavli Institute for Cosmology, Cambridge, where part of this work was completed. V.I. acknowledges support by the Kavli Foundation and the KICC Fellowship. S.K. is supported by a Flatiron Research
Fellowship at the Flatiron Institute, a division of the Simons Foundation and a Junior Research Fellowship from St Catharine’s College, Cambridge. S.K, M.A.B., and D.S. acknowledge support by European Research Council Starting Grant 638707 `Black holes and their host galaxies: coevolution across cosmic time'. M.A.B and D.S. additionally acknowledge support from the Science and Technology Facilities Council (STFC; grant number ST/W000997/1). M.A.B. acknowledges support from a UKRI Stephen Hawking Fellowship (EP/X04257X/1). This work made use of the following DiRAC facilities (www.dirac.ac.uk): the Data Analytic system at the University of Cambridge [funded by a BIS National E-infrastructure capital grant (ST/K001590/1), STFC capital grants ST/H008861/1 and ST/H00887X/1, and STFC DiRAC Operations grant ST/K00333X/1] and the COSMA Data Centric system at Durham University (funded by a BIS National E-infrastructure capital grant ST/K00042X/1, STFC capital grant ST/K00087X/1, DiRAC Operations grant ST/K003267/1 and Durham University). DiRAC is part of the National E-Infrastructure.
\section*{Data Availability}
Data employed in this manuscript will be shared upon reasonable request by contacting the corresponding author.\\

\noindent The raw Fable MPS data for each of the studied models at $z = \left[0, 0.4, 1.0, 2.0\right]$ is publicly available at:\\ 
\href{https://github.com/MartinAlvarezSergio/Fable_MPS}{https://github.com/MartinAlvarezSergio/Fable\_MPS}

\label{s:Public_Data}



\bibliographystyle{mnras}
\bibliography{fable_mps,references} 




\appendix

\section{Temporal Evolution of AGN Feedback Modes} \label{appsec:AGN_mode_evol}

\begin{figure}
    \centering
    \includegraphics[width=\columnwidth]{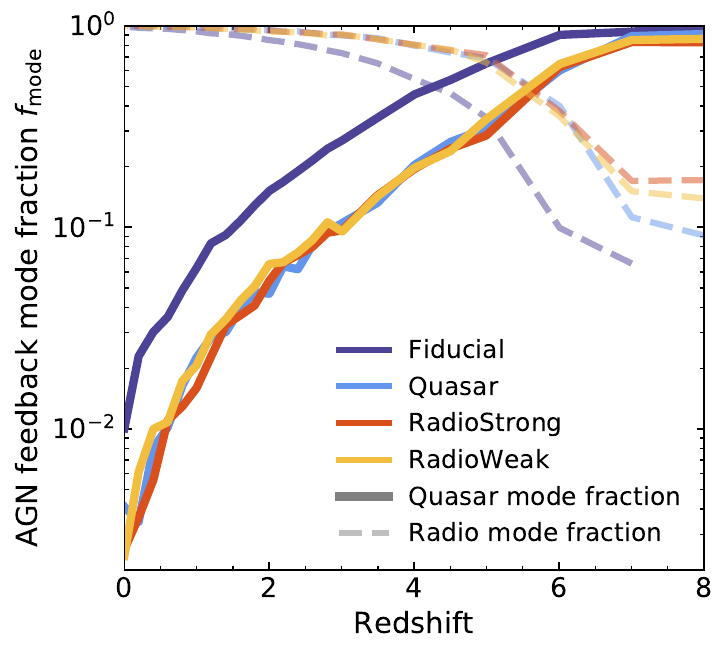}
    \caption{Quasar (thick solid lines) and radio (faint dashed lines) mode fractions for AGN in \fable~as a function of redshift. All models show a clear average decrease with redshift of AGN fraction in quasar mode. The quasar mode dominates at $z \gtrsim 4$, whereas the radio mode becomes more important below this redshift. The \Fiducial~model has an overall higher quasar mode fraction due to its lower quasar mode threshold (see Table~\ref{tab:sims_overview}).}
    \label{fig:AGNmodeRedshift}
\end{figure}

In the main body of this work, we found different AGN feedback models to exhibit distinct impacts on the MPS. Here we briefly discuss the evolution of AGN feedback modes (quasar vs radio) across the redshift evolution of \fable.

The fraction of AGN in each mode are shown in Figure \ref{fig:AGNmodeRedshift}, where solid lines represent the quasar mode and dashed lines denote the radio mode. The quasar mode is predominantly active at higher redshifts. This aligns with the expected behavior where quasar mode, being associated with high accretion rates, is more prevalent during the early, more chaotic epochs of galaxy formation. Conversely, the radio mode, which is often linked to maintenance feedback in more evolved systems, becomes dominant after $z \lesssim 4$. Despite a different feedback switch fraction in the \Fiducial~model, the measured impact on the MPS appears dominated by the combination of its higher radiative efficiency (akin to \RadioStrong) and the quasar duty cycle (also included in the \Quasar~simulation). At high redshift, the largest SMBHs are well within the quasar mode Eddington fraction regimen. Due to these AGN in these haloes being the main drivers of power suppression, we expect the feedback switch fraction to have relatively little effect on the MPS \footnote{The feedback switch mainly affects the AGN in the dwarf regime where there is a much weaker correlation between black hole mass and AGN activity, in particular at low redshifts \citep[see][]{Koudmani2021}}. Except for this aspect of the \Fiducial~model, all simulations exhibit a broadly similar evolution pattern. Such evolutionary trend for the AGN feedback modes provide further context for why our modifications to the quasar mode (\Quasar~and \Fiducial) are particularly important at high redshift ($z \gtrsim 1$), whereas the so-called `maintenance' radio mode effects reveal themselves after $z \lesssim 1$.

\section{Redshift Evolution of the Haloes MPS} \label{appsec:HalosEvolution}

\begin{figure*}
    \centering
    \includegraphics[width=2.08\columnwidth]{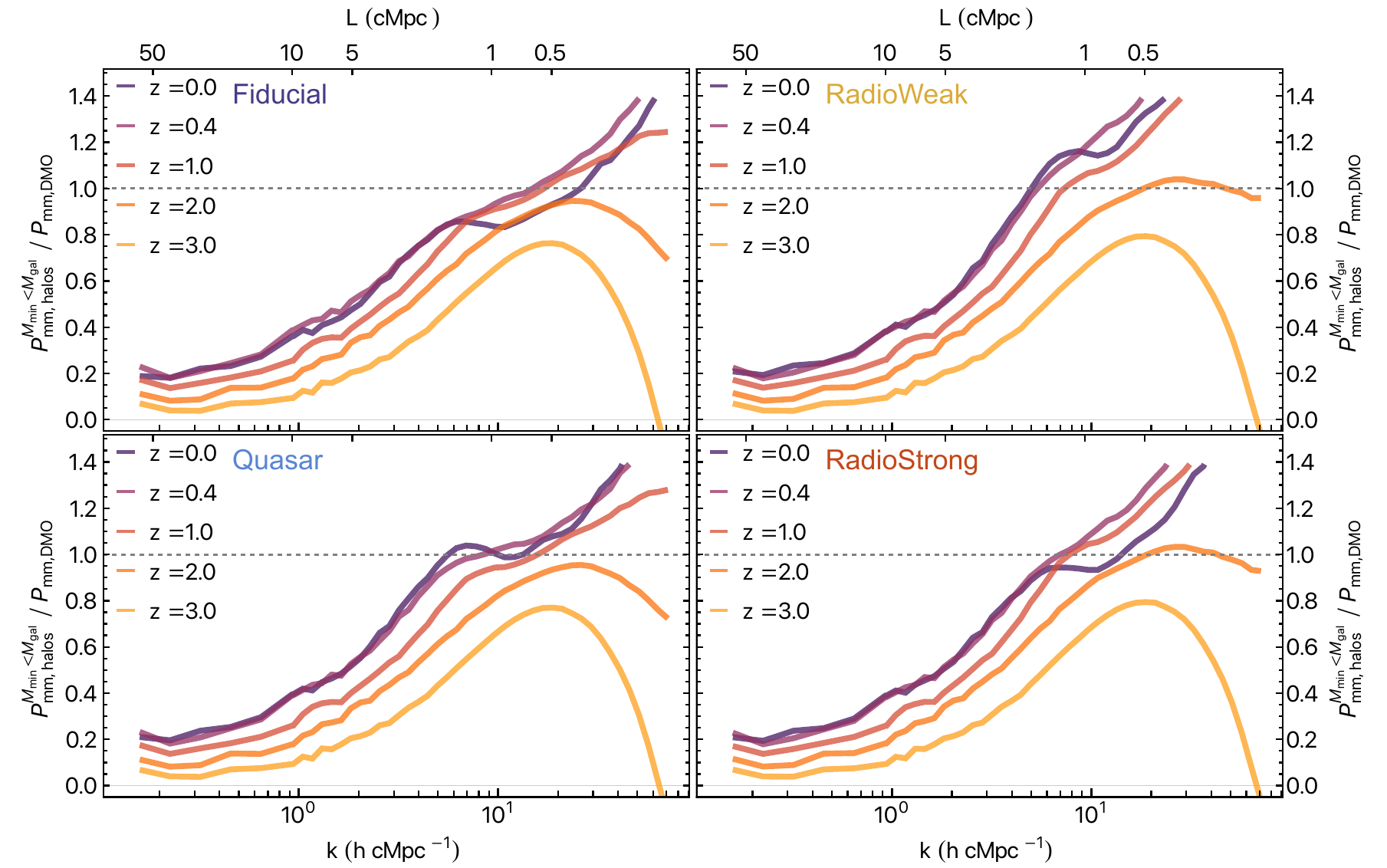}
    \caption{Redshift evolution of the fractional impact of baryonic physics on the MPS of all haloes in each simulation. Displayed quantities and simulations are the same as Figure~\ref{fig:redshiftEvolution}, except now we show $\Pmmhalo / \PDMO$ instead of $\Pmm / \PDMO$. Panels show the \Fiducial~(top left), \RadioWeak~(top right), \Quasar~(bottom left) and \RadioStrong~(bottom right) feedback models, respectively. The qualitative behaviour of the \Quasar~model being more efficient at higher redshift and the \RadioStrong~AGN more important at late times is also reproduced. The power of haloes has a subdominant contribution to $\Pmm$ at large scales ($k \sim 1\,\hcMpc$), and evolves approximately equivalently in all models.}
    \label{fig:HalosEvolution}
\end{figure*}

In Section~\ref{ss:AGNimpact} we studied how the total MPS evolved over redshift, whereas Sections~\ref{ss:GalaxiesCuts} and \ref{ss:HaloesTrace} addressed the effect of halo selection on the MPS, and the evolution of power at a fixed large ($\kscaleone$) and small ($\kscaleten$) scale, respectively. To complement these two aspects, here we provide further detail on the evolution of the haloes MPS. Understanding the evolution of the MPS for matter inside haloes will be particularly important as future observatories probe systems at $3 > z \gtrsim 0.5$ \citep{Ade2019}.

Figure~\ref{fig:HalosEvolution} is complementary to Figure~\ref{fig:redshiftEvolution}, but now shows the MPS from matter within haloes. It displays the redshift evolution of $\Pmmhalo$ from $z = 3$ to $z = 0$ in the \fable~AGN simulations: \Fiducial~(top left), \RadioWeak~(top right), \Quasar~(bottom left), and \RadioStrong~(bottom right). Overall, the trends observed in the main text regarding the temporal and scale-dependent impacts of different AGN feedback models are reproduced in the autopower of matter inside  haloes here. At larger scales ($k \lesssim 5 \hcMpc$), the amount of power residing within haloes is significantly smaller than the total $\Pmm$, but features a comparable evolution of all models. 

\bsp	
\label{lastpage}
\end{document}